\documentclass{sig-alternate-05-2015}

\usepackage[figure,vlined,linesnumbered]{algorithm2e}

\usepackage{float}
\usepackage{graphicx}
\usepackage{amsmath}

\usepackage{color}

\usepackage{fancyvrb}
\usepackage{verbatim}
\usepackage{multirow}
\usepackage{rotating}

\usepackage[T1]{fontenc}

\usepackage[scaled=0.90]{helvet}

\usepackage{mathpartir}
\usepackage[only,llbracket,rrbracket]{stmaryrd}

\usepackage[inference]{semantic}

\usepackage{textcomp}
\usepackage{balance}
\usepackage{url}

\renewcommand{\comment}[1]{}

\input{pygdef.tex}

\usepackage{etoolbox}
\makeatletter
\patchcmd{\maketitle}{\@copyrightspace}{}{}{}
\makeatother

\begin{document}

%\CopyrightYear{2016} 
%\setcopyright{acmcopyright}
%\conferenceinfo{ICSE '16,'}{May 14-22, 2016, Austin, TX, USA}
%\isbn{978-1-4503-3900-1/16/05}\acmPrice{\$15.00}
%\doi{http://dx.doi.org/10.1145/2884781.2884872}

\toappear{}

%\titlebanner{banner above paper title}        % These are ignored unless
%\preprintfooter{short description of paper}   % 'preprint' option specified.

\title{An Analysis of the Search Spaces for Generate and Validate Patch Generation Systems}

\author{
\alignauthor
\vspace{-9mm} Fan Long and Martin Rinard \\
       \affaddr{MIT EECS \& CSAIL, USA}\\
       \email{\{fanl, rinard\}@csail.mit.edu}
}

\maketitle

\begin{abstract}

We present the first systematic analysis of the characteristics of 
patch search spaces for automatic patch generation systems. We analyze the search
spaces of two current state-of-the-art systems, SPR and Prophet, with
16 different search space configurations. Our results are derived from an
analysis of 1104 different search spaces and 768 patch generation executions.
Together these experiments consumed over 9000 hours of CPU time
on Amazon EC2.

The analysis shows that 1) correct patches are sparse
in the search spaces (typically at most one correct
patch per search space per defect), 2) incorrect patches that
nevertheless pass all of the test cases in the validation test suite 
are typically orders of magnitude more abundant, and 3) 
leveraging information other than the test suite is therefore 
critical for enabling the system to successfully isolate correct patches. 

We also characterize a key tradeoff in the structure of the search
spaces. Larger and richer search spaces that contain correct patches for
more defects can actually cause systems to find fewer, 
not more, correct patches. We identify two reasons
for this phenomenon: 1) increased validation times 
because of the presence of more candidate patches and 2) 
more incorrect patches that pass the test suite and block the discovery of correct patches. 
These fundamental properties, which are all characterized for the first time in this
paper, help explain why past systems often 
fail to generate correct patches and help identify challenges, opportunities,
and productive future directions for the field.

%We present {\name}, a new program repair system that combines
%{\em staged program repair} and {\em condition synthesis}.
%These techniques enable SPR to work productively with a
%set of {\em parameterized transformation schemas} to 
%generate and efficiently search a rich space of program
%repairs.  Together these techniques enable {\name} to generate correct repairs
%for over five times as many defects as previous systems evaluated on the same
%benchmark set. 
\end{abstract}


\section{Introduction}

Software defects are a prominent problem in software development efforts. 
Motivated by the prospect of reducing human developer involvement, 
researchers have developed a range of techniques that are designed
to automatically correct defects. In this paper we focus on 
generate and validate patch generation systems, which work with a test
suite of test cases, generate a set of candidate patches, then test
the patched programs against the test suite to find a patch
that validates~\cite{Perkins:2009,LeGoues:2012, 
Weimer:2013, Qi:2014, Kim:2013, Kali, SPR, Prophet}.

\comment{
We identify 1) targeted techniques 
that repair specific classes of universal defects such as null
dereferences~\cite{Nulll}, out of bounds accesses~\cite{FOC}, and
infinite loops~\cite{Jolt,Bolt}, 2) synthesis-based techniques that
leverage formal specifications to produce patches that enable a 
defective program to satisfy the specification~\cite{SpecStuff}, 
and 3) generate-and-validate techniques, which work with a test
suite of inputs, generate a set of candidate patches, then test
the patched programs against the test suite to find a patch 
that validates~\cite{ClearView,GenProg,AE,RSRepair,PAR,Kali,SPR, Prophet}.
}

Patch quality is a key issue for generate and validate systems. 
Because the patches are validated only against the test cases in the 
test suite, there is no guarantee that the patch will enable 
the program to produce correct results for other test cases. Indeed,
recent research has shown that 1) the majority of patches accepted 
by many current generate and validate systems fail to generalize to 
produce correct results for test cases outside the validation test
suite~\cite{Kali,NOPOLTR,Overfitting} and 2) accepted patches
can have significant negative effects such as the introduction of
new integer and buffer overflow security vulnerabilities, 
undefined accesses, memory leaks, and the elimination of core application
functionality~\cite{Kali}. These negative effects highlight the
importance of generating not just {\em plausible patches} 
(we define plausible patches to be patches 
that pass all of the test cases in the patch validation
test suite) but {\em correct patches} that do not have latent
defects and do not introduce new defects or vulnerabilities. 

A rich search space that contains correct patches for target defects 
can be critical to the success of any automatic patch generation
system.  Indeed, recent research indicates 
that impoverished search spaces that contain very few correct patches 
is one of the reasons for the poor performance of some prominent
previous patch generation systems~\cite{LeGoues:2012, Weimer:2013, Kali}. While 
more recent systems work with patch spaces that contain significantly
more correct patches~\cite{SPR, Prophet}, continued progress in the
field requires even richer patch spaces that contain
more successful correct patches. But these richer spaces may also complicate the ability of the system
to identify correct patches within the larger sets of plausible
but incorrect patches. 

\subsection{SPR and Prophet}

SPR~\cite{SPR} and Prophet~\cite{Prophet} are two current state-of-the-art 
generate and validate patch generation systems.
Both systems work with a set of 
candidate program statements identified by an error localization 
algorithm. Both systems apply transformations to the statements
to obtain patches that they validate against a test
suite.  SPR uses a set of hand-coded patch prioritization heuristics. 
Prophet uses machine learning to characterize features of previously
successful human patches and prioritizes candidate patches according
to these features. The goal is to prioritize a correct patch 
as the first patch to validate.

The baseline SPR and Prophet search spaces contain correct patches
for 19 out of 69 defects in a benchmark set of defects from eight
open source projects~\cite{SPR, Prophet}.\footnote{
The paper that presented this benchmark set states that the benchmark
set contains 105 defects. An examination of the relevant commit logs
and applications indicates that 36 of these defects are actually 
deliberate functionality changes, not defects. In this paper we
focus on the remaining 69 actual defects as within the scope of the
paper. 
}
For 11 of these defects, the first SPR patch to validate is a correct patch. For
15 of these defects, the first Prophet patch to validate is a correct patch. 
The GenProg~\cite{LeGoues:2012}, AE~\cite{Weimer:2013}, RSRepair~\cite{Qi:2014}, 
and Kali~\cite{Kali} systems, in contrast, produce correct
patches for only 1 (GenProg, RSRepair) or 2 (AE, Kali) of the 
defects in this benchmark suite~\cite{Kali}. Moreover, the correct SPR
and Prophet patches for the remaining defects lie outside the GenProg,
RSRepair, and AE patch spaces, which suggests that these systems 
will {\em never} be able to produce correct patches for these
remaining defects. We note that this
benchmark set was not developed by us, but by the developers of
GenProg in an attempt to obtain a large, unbiased, and realistic
benchmark set~\cite{LeGoues:2012}.

As these results highlight, the characteristics of the
patch search space are central to the success of the patch generation system.
But despite the importance of the search space in the overall
success of the patch generation system, we have
been able to find no previous systematic investigation of 
how the structure of the patch search space influences critical characteristics
such as the density of correct and plausible patches and the
ability of the system to identify correct patches within the
broader class of plausible (but potentially incorrect) patches. 
Given the demonstrated negative effects of plausible but 
incorrect patches generated by previous systems~\cite{Kali}, 
these characteristics play a critical role in the overall 
evaluation (and eventual success or failure) of any patch generation 
system that may generate plausible patches with such negative effects. 

\subsection{Search Space Analysis} 

We present a systematic analysis of the SPR and Prophet search spaces. 
This analysis focuses on the density of correct and plausible patches
in the search spaces, on the ability of SPR and Prophet to prioritize
correct patches, and on the consequences of two 
kinds of changes, both of which increase the size of the search space:
1) increases in the number of candidate program statements 
to which patches are applied and 
2) new program transformation operators that can generate additional
patches. 
Starting from the SPR and Prophet
baseline search spaces, these changes make it possible to construct a collection
of search spaces, with each search space characterized by 
a combination of the set of transformations and the number of candidate 
program statements that together generate the search space. 

We perform our analysis on a benchmark set of 69 real world defects in 
eight open source large applications. This benchmark set 
was used to evaluate many
previous patch generation systems~\cite{LeGoues:2012, Weimer:2013, Kali, Qi:2014, SPR, Prophet}.
For each defect, we first analyze the full search space to determine whether or not
it contains a correct patch.  We acknowledge that, in general, determining whether a specific
patch corrects a specific defect can be difficult (or in some
cases not even well defined). But for the defects in the benchmark
set, this never happens --- the correct patches that are within
the search spaces are all small, all match the corresponding 
developer patches, and the distinction between correct and
incorrect patches is clear. 

Because the full search spaces can, in general, be too large to exhaustively search
within any reasonable time limit, we also consider the subset of the search spaces
that can be explored by SPR and Prophet within a reasonable timeout, in this
paper 12 hours. Working with these explored subsets of spaces, we analyze the
number of plausible and correct patches that each subset contains and the
effect of the SPR and Prophet patch prioritization on the ability of these
systems to identify correct patches within the much larger sets of plausible
but incorrect patches in these search spaces. 

\subsection{Results Overview}

Our experimental results indicate that:
\begin{itemize} % \itemsep 0pt \parskip 0pt 
\item {\bf Sparse Correct Patches:} Correct patches occur only sparsely within
the search spaces. For 45 of the 69 defects, the search spaces contain no
correct patches. For 15 of the remaining 24 defects, the search spaces contain
at most 1 correct patch. The largest number of correct patches for any defect
in any search space is 4.

\item {\bf Relatively Abundant Plausible Patches:} In comparison
with correct patches, plausible (but incorrect) patches are relatively abundant. 
For all of the benchmarks except php, the explored search spaces 
typically contain hundreds up to a thousand times more 
plausible patches than correct patches. 
These numbers highlight the difficulty of isolating correct patches among 
the large sets of plausible but incorrect patches.

The explored search spaces for php, in contrast, typically contain only
tens of times more plausible patches than correct patches. And for three
of the php defects, all of the (one or two) plausible patches are correct. 
The density of plausible patches is obviously related to the strength of 
the validation test suite --- weak test suites filter fewer incorrect patches. 
We attribute the difference in plausible
patch density between php and the other benchmarks to the strength of the 
php test suite --- the php test suite contains an order of magnitude more 
test cases than any other benchmark.

\item {\bf SPR and Prophet Effectiveness:} The SPR and Prophet patch
prioritization mechanisms are both effective at isolating correct patches
within the explored plausible patches. Despite the relatively scarcity of
correct patches, with the baseline search space, correct patches for 14 defects
are within the first ten patches to validate for SPR; correct patches for 16
defects are within the first ten patches to validate for Prophet.

%the probability of
%finding a correct patch within the first ten patches to validate
%(for a randomly chosen defect with a correct patch in the full search space) 
%ranges (depending on the search space) from 29\% to 58\% for SPR and from 33\% to 67\% for Prophet.

\item {\bf Search Space Tradeoffs:} Increasing the search space beyond
the SPR and Prophet baseline increases the number of defects that
have a correct patch in the search space. But it does not increase the ability of SPR
and Prophet to find correct patches for more defects --- in fact, these increases 
often cause SPR and Prophet to find correct patches for {\em fewer} defects!

We attribute this phenomenon to the following tradeoff. Increasing the search
space also increases the number of candidate patches and may increase the
number of plausible patches. The increased number of candidate patches consumes
patch evaluation time and reduces the density of the correct patches in the
search space. The increased number of plausible but incorrect patches increases
the chance that such patches will block the correct patch (i.e., that the system
will encounter plausible but incorrect patches as the first patches to validate). 

\end{itemize}

These facts highlight the importance of including information other
than the test suite in the patch evaluation process. SPR includes
information in the form of hand-coded patch prioritization 
heuristics. Prophet leverages information available via machine
learning from successful human patches. This information is
responsible for the ability of these systems to successfully identify correct patches
in the baseline search space. The results also highlight that
there is still room for improvement, especially with richer
search spaces that contain correct patches for more defects. 

\noindent{\bf Previous Systems:}
These facts also help explain past results from other systems.
GenProg, AE, and RSRepair generate very few correct patches~\cite{Kali}.
Part of the explanation is that the search space exploration
algorithms for these systems are no better than random
search~\cite{Qi:2014}.\footnote{
Other parts of the explanation include search spaces that 
apparently contain very few correct patches and errors
in the patch infrastructure that cause the system to accept
patches that do not even pass the test cases in the test
suites used to validate the patches~\cite{Kali}.
}
Once one appreciates the relative abundance of 
plausible but incorrect patches and the relative scarcity
of correct patches, it is clear that any algorithm that is
no better than random has very little chance of consistently
delivering correct patches without very strong test suites. 
And indeed, the majority of the patches from these
previous systems simply delete functionality and do not 
actually fix the defect~\cite{Kali}.

\noindent{\bf ClearView:} 
ClearView, in sharp contrast, does leverage
information other than the test suite, specifically learned invariants
from previous successful executions~\cite{Perkins:2009}. The results 
indicate a dramatic improvement in ClearView's ability to locate
patches that eliminate security vulnerability defects. For 
nine of the ten defects on which ClearView was evaluated, ClearView
successfully patched the defect after evaluating at most 
three candidate patches. These results highlight how targeting 
a defect class and leveraging fruitful sources of information can
dramatically increase the successful patch density. 

\subsection{Future Directions}

Our results highlight the scalability challenges associated
with generalizing existing search spaces to include correct
patches for more defects. One obvious future direction, deployed successfully
in past systems~\cite{Rinard:2004,Long:2014,DemskyR06}, is to address scalability
issues by developing smaller, more precisely targeted search 
spaces for specific classes of errors. An alternative is to 
infer transformation operations from correct
human patches (instead using manually defined transformations). 
The goal is to obtain a search space that contains correct patches for
common classes of commonly occurring defects while still remaining tractable. 

More broadly, it is now clear that generate and validate systems must
exploit information beyond current validation test suites if they
are to successfully correct any but the most trivial classes of defects~\cite{Perkins:2009,Prophet,Kali}.
One prominent direction is to exploit existing correct code in large code repositories to obtain 
new correct patches, either via sophisticated machine learning techniques
that learn to recognize or even automatically generate correct code,
automatically transferring correct code (either within or between applications),
or even generalizing and combining multiple blocks of correct code throughout
the entire software ecosystem.  Encouraging initial progress has been made in all of these 
directions~\cite{Prophet,Sidiroglou-Douskos15,son2013fix}.
The current challenge is to obtain larger blocks of correct code that implement more
sophisticated patches for broader classes of defects. 

An orthogonal direction is to obtain stronger test suites or even 
explicit specifications that can more effectively filter incorrect patches. 
One potential approach is to observe correct input/output pairs 
to learn to recognize or even automatically generate
correct outputs for (potentially narrowly targeted) classes of inputs.
Another approach leverages the availability of multiple implementations of the same basic functionality
(for example, multiple image rendering applications) to recognize correct outputs. 
Combining either of these two capabilities with automatic input generation could enable
the automatic generation of much stronger test suites (with potential
applications far beyond automatic patch generation). Specification mining
may also deliver (potentially partial) specifications that can help filter
incorrect patches. 

\subsection{Contributions}

This paper makes the following contributions:
\begin{itemize} % \itemsep 0pt \parskip 0pt 
%\item {\bf Systematic Experiments: } It presents a set of systematic
%experiments of applying SPR and Prophet to real world defects with different
%search spaces, including the analysis of 1104 different search spaces for the
%69 benchmark defects (we consider 16 search spaces for each defect) and the
%execution of 768 patch generation runs for the 24 defects whose correct patches
%are inside any of the considered search spaces (we run SPR and Prophet for each
%of the 16 search spaces for each defect). These experiments take more than 9000
%CPU hours on Amazon EC2.

\item {\bf Patch Space Analysis:} It presents an analysis of the patch search
spaces of SPR and Prophet, including how these patch spaces respond to the
introduction of new transformation operators and increases in candidate 
program statements. The analysis characterizes:
\begin{itemize} % \itemsep 0pt \parskip 0pt 
\item {\bf Correct Patches:} The density at which correct patches occur
in the full search spaces and the explored space subsets. 

\item {\bf Plausible Patches:} The density at which plausible patches
occur in the explored space subsets. 

\item {\bf Patch Prioritization:} The effectiveness
of the SPR and Prophet patch prioritization mechanisms at
isolating the few correct patches within the much larger set
of plausible patches (the vast majority of which are not correct). 

\end{itemize}
This paper presents the first characterization of how correct and plausible patch
densities respond to increases in the size and sophistication of the search space. 

\item {\bf Tradeoff:} It identifies and presents results
that characterize a tradeoff
between the size and sophistication of the search space and the
ability of the patch generation system to identify correct patches.
To the best of our knowledge, this is the first characterization
of this tradeoff. 

\item {\bf Results: } It presents experimental results from SPR and Prophet
with different search spaces. These results are derived from an
analysis of 1104 different search spaces for the 69 benchmark defects (we
consider 16 search spaces for each defect) and 768 patch generation executions
for the 24 defects whose correct patches are inside any of the considered
search spaces (we run SPR and Prophet for each of the 16 search spaces for each
defect). Together, these executions consumed over 9000 hours of CPU time on
Amazon EC2 cluster. The results show:
\begin{itemize} % \itemsep 0pt \parskip 0pt 
\item {\bf Sparse Correct Patches:} Correct patches occur very sparsely
within the patch spaces, with typically no more than one correct
patch in the search space for a given defect. 

\item {\bf Relatively Abundant Plausible Patches:} 
Depending on the strength of the validation test suite,
plausible patches are
either two to three orders of magnitude or one order of magnitude
more abundant than correct patches. 

\item {\bf Patch Prioritization Effectiveness:} The SPR and Prophet
patch prioritization algorithms exhibit substantial effectiveness
at isolating correct patches within the large set of plausible patches
(most of which are incorrect). 

\item {\bf Challenges of Rich Search Spaces:} 
The challenges associated with 
successfully searching such spaces include increased testing
overhead and increased chance of encountering plausible
but incorrect patches that block the subsequent discovery
of correct patches. 
\end{itemize}

\item {\bf Implications:} It discusses implications of the
facts and perspective presented in this paper,
both for the interpretation of results
from previous patch generation systems and for the design of
future patch generation systems. Specifically:
\begin{itemize}
\item {\bf Previous Systems:} It discusses how 
results from previous GenProg, AE, and RSRepair 
systems indicate that 1) the search spaces of these systems
contain very few correct patches (many fewer than the SPR and Prophet
search spaces) and 2) the search spaces contain many more plausible
but incorrect patches than correct patches.

\item {\bf ClearView:} It discusses how ClearView leverages
additional information other than the validation test suite
(specifically invariant information learned from 
previous successful executions) to obtain a patch space 
very dense in successful patches.

\item {\bf Future Systems:} It discusses how the ability to 
identify correct patches within a much larger set of plausible
patches is critical to ability of future systems to work 
successfully with larger, richer search spaces that contain
correct patches for more defects. 

\end{itemize}
\end{itemize}

Progress in automatic
patch generation systems requires the development of new, larger,
and richer patch search spaces that contain correct patches
for larger classes of defects. This paper characterizes
several central properties of the search spaces of current state-of-the-art
automatic patch generation systems. It also characterizes how 
these systems respond to changes in the size and sophistication
of their search spaces. It therefore identifies future productive
directions for the field and provides a preview of the issues that
the field will have to address to develop systems that work productively
with more sophisticated search spaces. 

\comment{

\begin{quote}
\framebox{
\parbox{0.39\textwidth}{
\noindent{\bf RQ1:}
By how much do the proposed increases in the candidate patch search
space increase the number of plausible and correct patches in the 
space?
}
}
\end{quote}

\begin{quote}
\framebox{
\parbox{0.39\textwidth}{
\noindent{\bf RQ1:}
Will exploring larger search space enable SPR and Prophet to generate correct
patches for more defects?
}
}
\end{quote}

\begin{quote}
\framebox{
\parbox{0.39\textwidth}{
\noindent{\bf RQ1a:}
Will considering more candidate program points from the error localization 
results enable SPR and Prophet to generate correct patches for more defects?
}
}
\end{quote}

\begin{quote}
\framebox{

\parbox{0.39\textwidth}{
\noindent{\bf RQ1b:}
Will considering more mutation operations enable SPR and Prophet to
automatically generate correct patches for more defects?
}
}
\end{quote}

\begin{quote}
\framebox{
\parbox{0.39\textwidth}{
\noindent{\bf RQ2:}
What are the key challenges for SPR and Prophet to automatically identify and generate
correct patches in the enlarged search space?
}
}
\end{quote}

\begin{quote}
\framebox{
\parbox{0.39\textwidth}{
\noindent{\bf RQ3:}
Why a larger search space does not enable SPR or Prophet to generate 
correct repairs for more defects?
}
}
\end{quote}

\begin{quote}
\framebox{
\parbox{0.39\textwidth}{
\noindent{\bf RQ4:}
Will the additional information that Prophet learns from past successful human
patches enable Prophet to generate correct patches for more defects than SPR
with the enlarged search space?
}
}
\end{quote}

\begin{quote}
\framebox{
\parbox{0.39\textwidth}{
\noindent{\bf RQ4:}
Will the enlarged search space enable SPR and Prophet to generate correct
patches for more defects if there are stronger test suites and/or human
intervention to filter plausible but incorrect patches?
}
}
\end{quote}
}

\section{SPR and Prophet}
\label{sec:overview}

We next present an overview of the two automatic patch generation systems, 
SPR~\cite{SPR} and Prophet~\cite{Prophet}, whose search space
and search algorithms we analyze. 

\subsection{Design Overview}
\label{sec:overview:overview}

To apply SPR or Prophet to a defective program, the user provides the system
with the program to patch and a validation test suite. The test
suite contains 1) a set of negative test cases which the original program does
not pass (these test cases expose the defect in the program) and 2) a set of
positive test cases which the original program already passes (these test cases
prevent regression). The test cases include correct outputs
for every input (the negative test cases produce different incorrect outputs). 
The system generates patches for the program with the following steps:

\noindent \textbf{Error Localization:} The system first uses an error localizer
to identify a set of candidate program statements to modify. The error localizer
recompiles the given application with additional instrumentation. It inserts a
call back before each statement in the source code to record a positive counter
value as the timestamp of the statement execution. The error localizer then
invokes the recompiled application on all test cases and produces,
based on the recorded timestamp values, a prioritized
list of target statements to modify. 
The error localizer prioritizes statements that are 1)
executed with more negative test cases, 2) executed with fewer positive
test cases, and 3) executed later during executions with negative test
cases. See the SPR and Prophet papers for more
details~\cite{SPR,Prophet}.

\noindent \textbf{Apply Transformations:}
The system then applies a set of transformations to the identified program
statements to generate the search space of candidate patches. 
SPR and Prophet consider the following transformation schemas~\cite{SPR}: 
\begin{itemize} % \itemsep 0pt \parskip 0pt 
\item {\bf Condition Refinement:} Given a target if statement to patch, the system
transforms the condition of the if statement by conjoining or disjoining an
additional condition to the original if condition. The following two patterns
implement the transformation:
\vspace{-1mm}
\begin{verbatim}
if (C) { ... } => if (C && P) { ... }
if (C) { ... } => if (C || P) { ... }
\end{verbatim}
\vspace{-1mm}
Here \verb+if (C) { ... }+ is the target statement to patch
in the original program. \verb+C+ is the original condition that appears in the program.
\verb+P+ is a new condition produced by a condition
synthesis algorithm~\cite{SPR,Prophet}.

\item {\bf Condition Introduction:} Given a target statement, the system
transforms the program so that the statement executes only if a guard
condition is true. 
The following pattern implements the transformation:
\vspace{-1mm}
\begin{verbatim}
S => if (P) S
\end{verbatim}
\vspace{-1mm}
Here \verb+S+ is the target statement to patch in the original program and
\verb+P+ is a new synthesized condition.

\item {\bf Conditional Control Flow Introduction:} Before a target statement, 
the system inserts a new
control flow statement (return, break, or goto an existing label) that executes
only if a guard condition is true.
The following patterns implement the transformation:
\vspace{-1mm}
\begin{verbatim}
S => if (P) break; S
S => if (P) continue; S
S => if (P) goto L; S
\end{verbatim}
\vspace{-1mm}
Here \verb+S+ is the target statement to patch in the original program, 
\verb+P+ is a new synthesized condition, and 
\verb+L+ is an existing label in the procedure containing \verb+S+.

\item \textbf{Insert Initialization: } Before a target statement, the system
inserts a memory initialization statement.

\item \textbf{Value Replacement: } Given a target statement, 
replace an expression in the statement with another expression. 

\item \textbf{Copy and Replace: } Given a target statement, the system copies
an existing statement to the program point before the target statement and then
applies a Value Replacement transformation to the copied statement. 
\end{itemize}

\noindent{\bf Condition Synthesis:} 
The baseline versions of SPR and Prophet work with synthesized conditions
\verb+P+ of the form \verb+E == K+ and \verb+E != K+. Here
\verb+E+ is a {\em check expression}, which we define as
either a local variable, a global variable, or a sequence of structure field accesses. 
Each check expression \verb+E+ must appear in the basic block containing the synthesized condition. 
\verb+K+ is a {\em check constant}, which
we define as a constant drawn from the set of values that the check expression 
\verb+E+ takes on
during the instrumented executions of the unpatched program on the negative test cases. 

\noindent{\bf Value Replacement:} 
The baseline versions of SPR and Prophet replace either 1) one variable in the
target statement with another variable that appears in the basic block
containing the statement, 2) an invoked function in the statement with another
function that has compatible type signatures, or 3) a constant in the statement with another
constant that appears in the function containing the statement. 

\noindent \textbf{Evaluate Candidate Patches:} The system then evaluates
candidate patches in the search space against the supplied test cases. To
efficiently explore the search space, SPR and Prophet use {\em staged program repair}~\cite{SPR, Prophet}.
At the first stage, the system operates with
parameterized candidate patch templates, which may contain an abstract
expression. It instantiates and evaluates concrete patches from a template
only if the system determines that there may be a concrete patch from the
template that passes the test cases. 

For the first three transformation schemas (these schemas manipulate
conditions), the system first introduces an abstract condition into the program
and determines whether there is a sequence of branch directions for the abstract
condition that will enable the patched program to pass the test
cases. If so, the system then synthesizes concrete conditions to generate
patches. 

\subsection{Extensions}

%We extend the SPR and Prophet condition synthesis algorithm to include the
%``\texttt{<}'' and ``\texttt{>}'' operators and to also consider comparisons
%between two expressions. In the rest of this paper, we use ``CExt'' to denote
%this search space extension.
We implement three extensions to the SPR and Prophet search spaces:
considering more candidate program statements 
to patch, synthesizing more sophisticated conditions, 
and evaluating more complicated value replacement transformations.

\noindent \textbf{More Program Statements to Patch: } The baseline 
SPR and Prophet configurations consider the first 200 program statements identified by the 
error localizer. We modify SPR and Prophet to consider
the first 100, 200, 300, and 2000 statements.

\noindent \textbf{Condition Synthesis Extension (CExt):} 
We extend the baseline SPR and Prophet condition synthesis algorithm to include the
``\texttt{<}'' and ``\texttt{>}'' operators and to also consider comparisons
between two check expressions (e.g., $\texttt{E} \texttt{ < } \texttt{K}$, 
$\texttt{E}_1 \texttt{ == } \texttt{E}_2$, and 
$\texttt{E}_1 \texttt{ > } \texttt{E}_2$, where 
$\texttt{E}$, $\texttt{E}_1$, and $\texttt{E}_2$ are
check expressions 
and \verb+K+ is a check constant).  In the rest of this paper, we use ``CExt'' to denote
this search space extension.

\noindent \textbf{Value Replacement Extension (RExt): } 
We extend the baseline SPR and Prophet replacement transformations to also
replace a variable or a constant in the target statement with an expression
that is composed of either 1) a unary operator and an atomic value (i.e., a
variable or a constant) which appears in the basic block containing the
statement or 2) a binary operator and two such atomic values. The operators
that SPR and Prophet consider are ``\texttt{+}'', ``\texttt{-}'',
``\texttt{*}'', ``\texttt{==}'', ``\texttt{!=}'', and ``\texttt{\&}''. In the
rest of this paper, we use ``RExt'' to denote this search space extension.

\subsection{SPR Prioritization Order}

SPR uses a set of hand-coded heuristics to prioritize its search of the generated
patch space. These heuristics prioritize patches in the following order: 
1) patches that change only a branch condition (e.g., tighten and
loosen a condition),
2) patches that insert an if-statement before 
the first statement of a compound statement (i.e., C code block),
3) patches that insert an if-guard around a statement,
4) patches that replace a statement, insert an initialization statement,
insert an if-statement, or 
insert a statement before the first statement of a compound statement,
and 5) finally all the remaining patches.
For each kind of patch, it prioritizes statements to patch in the 
error localization order. 
%So SPR first tests all tighten patches on all
%target statements that the error localizer identifies, then all Loosen
%patches on all identified target statements, and so on.  

\subsection{Prophet Prioritization Order}

Prophet searches the same patch space as SPR, but works
with a corpus of correct patches from human developers. It 
processes this corpus to learn a probabilistic model that assigns a 
probability to each candidate patch in the search space. This probability 
indicates the likelihood that the patch is correct.  It then uses this model
to prioritize its search of the patch space. 

A key idea behind Prophet is that patch correctness depends
on not just the patch itself, but also on how the patch interacts with
the surrounding code:
\begin{itemize} % \itemsep 0pt \parskip 0pt 
\item {\bf Extract Features: } For each patch in the corpus, Prophet
analyzes a structural diff of the abstract syntax trees of the original and patched code to 
extract both 1) features which summarize how the patch modifies the program 
given characteristics of the surrounding code and 2) 
features which summarize relationships between roles that values accessed by the patch
play in the original unpatched program and in the patch. 

\sloppypar{
\item {\bf Learn Model Parameters:} Prophet operates with a parameterized
log-linear probabilistic model in which the model parameters can be interpreted as 
weights that capture the importance of different features. Prophet learns the model 
parameters via maximum likelihood
estimation, i.e., the Prophet learning algorithm attempts to find parameter values that maximize the
probability of observing the collected training database in the probabilistic model. 
}
\end{itemize}

Prophet uses the trained model to rank the patches according to its learned
model of patch correctness, then evaluates the patches in that order. Previous
results (as well as additional results presented in this paper) show that
this learned patch correctness model outperforms SPR's
heuristics~\cite{Prophet}. This result highlights how leveraging information
available in existing large software development projects can significantly
improve our ability to automatically manipulate large software systems.

\section{Study Methodology}
\label{sec:methodology}

\noindent We next present our study methodology. 
% Section~\ref{sec:methodology:extension}
% presents three search space extensions which we consider in our study.
% Section~\ref{sec:methodology:benchmark} presents the benchmark defects.
% Section~\ref{sec:methodology:setup} presents the experimental setup for our study.

\subsection{Benchmark Applications}
\label{sec:methodology:benchmark}

We use a benchmark set of 69 real world defects to
perform our search space study. Those defects are from eight large open source
applications, libtiff, lighttpd, the php interpreter, gmp, gzip, python,
wireshark, and fbc~\cite{LeGoues:2012}.
Note that the original benchmark set also includes 36 
ostensible defects which correspond to 
deliberate functionality changes, not defects,
during the application development~\cite{SPR}. We exclude
those functionality changes as outside the scope of our study because they
are not actual defects. 

\begin{table}
\begin{tabular}{|l|c|c|c|c|c|}
\hline
\textbf{App.} & \textbf{LoC} & \textbf{Tests} & \textbf{Defects} & \textbf{SPR} & \textbf{Prophet} \\
\hline
libtiff & 77k & 78 & 8 & 1/3 & 2/3 \\
lighttpd & 62k & 295 & 7 & 0/0 & 0/0 \\
php & 1046k & 8471 & 31 & 9/13 & 10/13 \\
gmp & 145k & 146 & 2 & 1/1 & 1/1 \\
gzip & 491k & 12 & 4 & 0/1 & 1/1 \\
python & 407k & 35 & 9 & 0/0 & 0/0 \\
wireshark & 2814k & 63 & 6 & 0/0 & 0/0 \\
fbc & 97k & 773 & 2 & 0/1 & 1/1 \\
\hline
Total & & & 69 & 11/19 & 15/19 \\
\hline
%\multirow{2}{*}{\textbf{LoC}} & \multirow{2}{*}{\textbf{Tests}} & 
%\multirow{2}{*}{\textbf{\parbox{1.4cm}{Defects/ Changes}}} & \multicolumn{4}{c|}{\textbf{Plausible}}  
%& \multicolumn{4}{c|}{\textbf{Correct}} \\
%libtiff & 77k & 78 & 8/16 & 5/0 & 5/0 & 3/0 & 5/0 & 2/0 & 1/0 & 0/0 & 0/0 \\
%lighttpd & 62k & 295 & 7/2 & 3/1 & 3/1 & 4/1 & 3/1 & 0/0 & 0/0 & 0/0 & 0/0 \\
%php & 1046k & 8471 & 31/13 & 16/2 & 15/2 & 5/0 & 7/0 & 10/0  & 9/0 & 1/0 & 2/0 \\
%gmp & 145k & 146 & 2/0 & 2/0 & 2/0 & 1/0 & 1/0 & 1/0 & 1/0 & 0/0 & 0/0 \\
%gzip & 491k & 12 & 4/1 & 2/0 & 2/0 & 1/0 & 2/0 & 1/0 & 0/0 & 0/0 & 0/0 \\
%python & 407k & 35 & 9/2 & 5/1 & 5/1 & 0/1 & 2/1 & 0/0 & 0/0 & 0/1 & 0/1 \\
%wireshark & 2814k & 63 & 6/1 & 4/0 & 4/0 & 1/0 & 4/0 & 0/0 & 0/0 & 0/0 & 0/0 \\
%fbc & 97k & 773 & 2/1 & 1/0 & 1/0 & 1/0 & 1/0 & 1/0 & 0/0 & 0/0 & 0/0 \\
%\hline
%Total & & & 69/36 & 38/4 & 37/4 & 16/2 & 25/2 & 15/0 & 11/0 & 1/1 & 2/1 \\
%\hline
\end{tabular}

\vspace{-1mm}
\caption{Benchmark Applications}
\label{tab:benchmark}
\vspace{-1mm}
\end{table}

Table~\ref{tab:benchmark} summarizes our benchmark defects. The first column
(App.) presents the name of each application. The second column (LoC) presents
the number of lines of code in the application. The third column (Tests) presents the
number of the test cases in the supplied test suite of the application. php is the
outlier, with an order of magnitude more test cases than any other application.
The fourth column (Defects) presents the number of defects 
in the benchmark set for each application. 

The fifth column (SPR) and the sixth column (Prophet) present the patch
generation results for the baseline versions of SPR~\cite{SPR} and
Prophet~\cite{Prophet}, respectively. Each entry is of the form ``X/Y'', where
Y is the number of defects whose correct patches are inside the search
space, while X is the number of defects for which the system automatically
generates a correct patch as the first generated patch. 

\comment{
SPR and Prophet explore the same baseline search space, which contains correct patches
for 19 defects. With the prioritization order determined by the hand-coded
heuristics, SPR finds the correct patch as the first patch to 
validate for 11 out of the 19 defects.
Prophet's probabilistic patch correctness model (learned from past
successful human patches) enables Prophet to find the correct patch
as the first patch to validate for 15 out of the 19 defects. 
}

\begin{table}
\begin{tabular}{|l|c|c|c|}
\hline
\multirow{2}{*}{\textbf{Defect}} & \textbf{Localization} & & \\
 & \textbf{Rank} & \textbf{RExt} & \textbf{CExt} \\
\hline
lighttpd-2661-2662 & 1926 & No & No \\ 
\hline
lighttpd-1913-1914 & 280 & No & Yes \\
\hline
python-70056-70059 & 214 & No & Yes \\
\hline
python-69934-69935 & 136 & Yes & No \\
\hline
gmp-14166-14167 & 226 & Yes & No \\
\hline
\end{tabular}

\caption{Search Space Extensions}
\label{tab:extension}
\vspace{-2mm}
\end{table}

\begin{table*}[t]
\small
\begin{tabular}{|l|c|c|c|c|c|c|c|c|c|c|}
\hline
\multirow{2}{*}{\bf System} & {\bf Loc.} & {\bf Space} & \multicolumn{2}{|c|}{\bf Correct} & {\bf Plausible} & \multirow{2}{*}{\bf Timeout} & {\bf Space} & {\bf Correct} & {\bf Plausible} & {\bf Correct} \\
\cline{4-5}
    & {\bf Limit} & {\bf Extension} & {\bf In Space} & {\bf First} & \& {\bf Blocked} &                         & {\bf Size} & {\bf Rank} & {\bf in 12h}   &  {\bf in 12h} \\
\hline
SPR &  100 &  No &  4 &  1 &  7(3) &  3(0) &  20068.5 &  4614.0 &  8(2747) &  4(5) \\
\hline
SPR &  100 &  CExt &  4 &  1 &  7(3) &  3(0) &  20068.5 &  4614.0 &  8(11438) &  3(4) \\
\hline
SPR &  100 &  RExt &  4 &  1 &  7(3) &  3(0) &  21999.8 &  6004.8 &  8(2742) &  4(5) \\
\hline
SPR &  100 &  RExt+CExt &  4 &  1 &  7(3) &  3(0) &  21999.8 &  6004.8 &  8(11192) &  3(4) \\
\hline
SPR &  200 &  No &  6 &  2 &  7(4) &  2(0) &  46377.6 &  17889.5 &  9(2558) &  6(8) \\
\hline
SPR &  200 &  CExt &  6 &  2 &  7(4) &  2(0) &  46377.6 &  17889.5 &  9(10823) &  4(6) \\
\hline
SPR &  200 &  RExt &  7 &  2 &  7(4) &  2(1) &  52864.3 &  24759.9 &  9(3753) &  6(8) \\
\hline
SPR &  200 &  RExt+CExt &  7 &  2 &  7(4) &  2(1) &  52864.3 &  24759.9 &  9(10855) &  4(6) \\
\hline
SPR &  300 &  No &  6 &  1 &  8(5) &  2(0) &  73559.6 &  22960.0 &  9(2818) &  6(8) \\
\hline
SPR &  300 &  CExt &  8 &  1 &  8(6) &  2(1) &  73559.6 &  30761.8 &  9(10237) &  4(6) \\
\hline
SPR &  300 &  RExt &  8 &  1 &  8(6) &  2(1) &  82187.2 &  32951.4 &  9(2069) &  7(8) \\
\hline
SPR &  300 &  RExt+CExt &  10 &  1 &  8(7) &  2(2) &  82187.2 &  37427.4 &  9(10455) &  5(6) \\
\hline
SPR &  2000 &  No &  7 &  2 &  7(5) &  2(0) &  523753.8 &  157038.4 &  9(751) &  5(6) \\
\hline
SPR &  2000 &  CExt &  9 &  2 &  7(6) &  2(1) &  523753.8 &  156495.1 &  9(6123) &  4(5) \\
\hline
SPR &  2000 &  RExt &  9 &  2 &  7(6) &  2(1) &  574325.1 &  200996.7 &  9(657) &  5(6) \\
\hline
SPR &  2000 &  RExt+CExt &  11 &  2 &  7(7) &  2(2) &  574325.1 &  192034.0 &  9(5831) &  4(5) \\
\hline
Prophet &  100 &  No &  4 &  4 &  4(0) &  3(0) &  20068.5 &  589.2 &  8(2481) &  4(5) \\
\hline
Prophet &  100 &  CExt &  4 &  3 &  5(1) &  3(0) &  20068.5 &  589.2 &  8(11901) &  3(4) \\
\hline
Prophet &  100 &  RExt &  4 &  4 &  4(0) &  3(0) &  21999.8 &  520.5 &  8(2183) &  4(5) \\
\hline
Prophet &  100 &  RExt+CExt &  4 &  3 &  5(1) &  3(0) &  21999.8 &  520.5 &  8(11595) &  3(4) \\
\hline
Prophet &  200 &  No &  6 &  5 &  4(1) &  2(0) &  46377.6 &  11382.8 &  9(2564) &  5(7) \\
\hline
Prophet &  200 &  CExt &  6 &  4 &  5(2) &  2(0) &  46377.6 &  11382.8 &  9(10968) &  4(6) \\
\hline
Prophet &  200 &  RExt &  7 &  5 &  4(1) &  2(1) &  52864.3 &  19581.0 &  9(1939) &  5(6) \\
\hline
Prophet &  200 &  RExt+CExt &  7 &  4 &  5(2) &  2(1) &  52864.3 &  19581.0 &  9(10928) &  4(5) \\
\hline
Prophet &  300 &  No &  6 &  4 &  5(2) &  2(0) &  73559.6 &  11997.2 &  9(2555) &  5(7) \\
\hline
Prophet &  300 &  CExt &  8 &  3 &  6(4) &  2(1) &  73559.6 &  14466.8 &  9(10948) &  4(6) \\
\hline
Prophet &  300 &  RExt &  8 &  4 &  5(3) &  2(1) &  82187.2 &  25769.1 &  9(1548) &  5(6) \\
\hline
Prophet &  300 &  RExt+CExt &  10 &  3 &  6(5) &  2(2) &  82187.2 &  25455.1 &  9(10886) &  4(5) \\
\hline
Prophet &  2000 &  No &  7 &  4 &  5(3) &  2(0) &  523753.8 &  188588.4 &  9(1229) &  5(7) \\
\hline
Prophet &  2000 &  CExt &  9 &  3 &  6(5) &  2(1) &  523753.8 &  156555.8 &  9(8208) &  4(6) \\
\hline
Prophet &  2000 &  RExt &  9 &  3 &  6(5) &  2(1) &  574325.1 &  170715.4 &  9(1216) &  5(6) \\
\hline
Prophet &  2000 &  RExt+CExt &  11 &  2 &  7(7) &  2(2) &  574325.1 &  148288.4 &  9(7919) &  4(5) \\
\hline
\end{tabular}

\caption{Patch Generation Results with Search Space Extensions (excluding php)}
\label{tab:results-nophp}
\vspace{-2mm}
\end{table*}

\subsection{Experimental Setup}
\label{sec:methodology:setup}

\noindent \textbf{Configure Systems: } 
We run SPR and Prophet on each of the 16 different search space configurations
derived from all possible combinations of 1) working with the first 100, 200,
300, or 2000 program statements identified by the error localizer, 2) whether
to enable value replacement extension (RExt), and 3) whether to enable
condition synthesis extension (CExt).

We run all of our experiments except those of fbc on Amazon EC2 Intel Xeon
2.6GHz machines running Ubuntu-64bit server 14.04. fbc runs only in 32-bit
environments, so we run all fbc experiments on EC2 Intel Xeon 2.4GHz machines
running Ubuntu-32bit 14.04.

\noindent \textbf{Generate Search Spaces: } For each search space
and each of the 69 defects in the benchmark set, 
we run the configured SPR and Prophet to generate and print the
search space for that defect.  We then analyze the
generated search space and determine whether 
the space contains a correct patch for the defect. 

With all three search space extensions, the generated SPR and Prophet search
spaces contain correct patches for five more defects (i.e., 24 defects in
total) than the baseline search space. Table~\ref{tab:extension} summarizes
these five defects. The first column (Defect) contains entries of the form
X-Y-Z, where X is the name of the application that contains the defect, Y is
the defective revision in the application repository, and Z is the reference
fixed revision in the repository. The second column (Localization Rank)
presents the error localization rank of the modified program statement in the
correct patch for the defect. The third column (RExt) presents whether the
correct patches for the defect require the RExt extension (value replacement extension) 
The fourth column (CExt) presents whether the correct
patches for the defect require the CExt extension (condition synthesis extension).

\noindent \textbf{Generate Patches: } For each search space and each
of the 24 defects with correct patches in the search space, we
run SPR and Prophet to explore the search space for the defect. 
For each run, we record all of the plausible patches that the system
discovers within the 12 hour timeout. 

\comment{
patches generated by the system for the defect within
12 hours. 

There are three possible outcomes:
\begin{itemize}
\item {\bf Correct Patch:} The system successfully generates a correct patch
that removes the defect in the program without introducing regression. The
correct patch is the first patch generated by the run.

\item {\bf Plausible but Incorrect Patch:} The system generates a plausible but
incorrect patch.  If the run continues, it may generate a correct patch
later. In this situation, we do not count the system as automatically generating
the correct patch because the system requires external help to filter those
incorrect patches generated before the correct patch.

\item {\bf Timeout:} The system fails to generate a patch that passes the
supplied test cases within the time limit. In our experiments, the time limit
for each run is 12 hours. We terminate the run if it does not finish within 12
hours.
\end{itemize}
}

\noindent \textbf{Analyze Patches: } For each defect, we analyze the
generated plausible patches for the defect to determine whether the patch is correct
or incorrect.

\begin{table*}[t]
\small
\begin{tabular}{|l|c|c|c|c|c|c|c|c|c|c|}
\hline
\multirow{2}{*}{\bf System} & {\bf Loc.} & {\bf Space} & \multicolumn{2}{|c|}{\bf Correct} & {\bf Plausible} & \multirow{2}{*}{\bf Timeout} & {\bf Space} & {\bf Correct} & {\bf Plausible} & {\bf Correct} \\
\cline{4-5}
    & {\bf Limit} & {\bf Extension} & {\bf In Space} & {\bf First} & \& {\bf Blocked} &                         & {\bf Size} & {\bf Rank} & {\bf in 12h}   &  {\bf in 12h} \\
\hline
SPR &  100 &  No &  12 &  8 &  3(3) &  2(1) &  13446.5 &  4157.8 &  11(237) &  9(14) \\
\hline
SPR &  100 &  CExt &  12 &  8 &  4(4) &  1(0) &  13446.5 &  4157.8 &  12(415) &  10(12) \\
\hline
SPR &  100 &  RExt &  12 &  8 &  4(4) &  1(0) &  14026.7 &  4360.8 &  12(288) &  11(15) \\
\hline
SPR &  100 &  RExt+CExt &  12 &  8 &  4(4) &  1(0) &  14026.7 &  4360.8 &  12(421) &  10(12) \\
\hline
SPR &  200 &  No &  13 &  9 &  3(3) &  1(1) &  26512.0 &  7369.8 &  12(197) &  10(15) \\
\hline
SPR &  200 &  CExt &  13 &  9 &  3(3) &  1(1) &  26512.0 &  7369.8 &  12(330) &  10(13) \\
\hline
SPR &  200 &  RExt &  13 &  9 &  3(3) &  1(1) &  28158.2 &  7984.5 &  12(200) &  10(15) \\
\hline
SPR &  200 &  RExt+CExt &  13 &  9 &  3(3) &  1(1) &  28158.2 &  7984.5 &  12(323) &  10(13) \\
\hline
SPR &  300 &  No &  13 &  8 &  4(4) &  1(1) &  41859.8 &  10915.2 &  12(176) &  9(14) \\
\hline
SPR &  300 &  CExt &  13 &  8 &  4(4) &  1(1) &  41859.8 &  10915.2 &  12(305) &  9(12) \\
\hline
SPR &  300 &  RExt &  13 &  8 &  4(4) &  1(1) &  44631.1 &  12440.9 &  12(179) &  9(14) \\
\hline
SPR &  300 &  RExt+CExt &  13 &  8 &  4(4) &  1(1) &  44631.1 &  12440.9 &  12(313) &  9(12) \\
\hline
SPR &  2000 &  No &  13 &  5 &  2(2) &  6(6) &  327905.6 &  81570.5 &  7(58) &  5(6) \\
\hline
SPR &  2000 &  CExt &  13 &  5 &  2(2) &  6(6) &  327905.6 &  81570.5 &  7(126) &  5(6) \\
\hline
SPR &  2000 &  RExt &  13 &  5 &  3(3) &  5(5) &  356104.8 &  83997.9 &  8(59) &  5(6) \\
\hline
SPR &  2000 &  RExt+CExt &  13 &  5 &  2(2) &  6(6) &  356104.8 &  83997.9 &  7(127) &  5(6) \\
\hline
Prophet &  100 &  No &  12 &  8 &  3(3) &  2(1) &  13446.5 &  2599.4 &  11(279) &  11(15) \\
\hline
Prophet &  100 &  CExt &  12 &  6 &  6(6) &  1(0) &  13446.5 &  2599.4 &  12(466) &  11(11) \\
\hline
Prophet &  100 &  RExt &  12 &  9 &  3(3) &  1(0) &  14026.7 &  3433.8 &  12(327) &  11(15) \\
\hline
Prophet &  100 &  RExt+CExt &  12 &  6 &  6(6) &  1(0) &  14026.7 &  3433.8 &  12(458) &  11(11) \\
\hline
Prophet &  200 &  No &  13 &  10 &  3(3) &  0(0) &  26512.0 &  3522.1 &  13(285) &  13(18) \\
\hline
Prophet &  200 &  CExt &  13 &  7 &  6(6) &  0(0) &  26512.0 &  3522.1 &  13(447) &  12(13) \\
\hline
Prophet &  200 &  RExt &  13 &  10 &  3(3) &  0(0) &  28158.2 &  4504.4 &  13(296) &  12(17) \\
\hline
Prophet &  200 &  RExt+CExt &  13 &  7 &  6(6) &  0(0) &  28158.2 &  4504.4 &  13(434) &  12(13) \\
\hline
Prophet &  300 &  No &  13 &  10 &  3(3) &  0(0) &  41859.8 &  4319.6 &  13(280) &  13(18) \\
\hline
Prophet &  300 &  CExt &  13 &  7 &  6(6) &  0(0) &  41859.8 &  4319.6 &  13(425) &  12(13) \\
\hline
Prophet &  300 &  RExt &  13 &  10 &  3(3) &  0(0) &  44631.1 &  5403.1 &  13(283) &  12(17) \\
\hline
Prophet &  300 &  RExt+CExt &  13 &  7 &  6(6) &  0(0) &  44631.1 &  5403.1 &  13(422) &  12(13) \\
\hline
Prophet &  2000 &  No &  13 &  7 &  2(2) &  4(4) &  327905.6 &  21118.6 &  9(117) &  7(10) \\
\hline
Prophet &  2000 &  CExt &  13 &  4 &  4(4) &  5(5) &  327905.6 &  21118.6 &  8(153) &  6(6) \\
\hline
Prophet &  2000 &  RExt &  13 &  6 &  2(2) &  5(5) &  356104.8 &  25168.5 &  8(104) &  6(9) \\
\hline
Prophet &  2000 &  RExt+CExt &  13 &  4 &  4(4) &  5(5) &  356104.8 &  25168.5 &  8(183) &  6(6) \\
\hline
\end{tabular}

\caption{Patch Generation Results with Search Space Extensions (php only)}
\label{tab:results-php}
\vspace{-2mm}
\end{table*}

\section{Experimental Results}
\label{sec:results}

We next present the experimental results. php is an outlier with a test suite
that contains an order of magnitude more test cases than the other
applications. We therefore separate the php results from the results from other
benchmarks. We present the result summary for all of the 24 defects for which
any of the search spaces contains a correct patch. See Appendix for the
detailed results of each defect in all different search space configurations.

\comment{
\begin{table*}
\small
\begin{tabular}{|l|c|c|c|c|c|c|c|c|c|c|}
\hline
\multirow{2}{*}{\bf System} & {\bf Loc.} & {\bf Space} & \multicolumn{2}{|c|}{\bf Correct} & {\bf Plausible} & \multirow{2}{*}{\bf Timeout} & {\bf Space} & {\bf Correct} & {\bf Plausible} & {\bf Correct} \\
                        & {\bf Limit} & {\bf Extension} & {\bf In Space} & {\bf First} & \& {\bf Blocked} &                         & {\bf Size} & {\bf Rank} & {\bf in 12h}   &  {\bf in 12h} \\
\hline
SPR &  100 &  No &  16 &  9 &  10(6) &  5(1) &  33515.3 &  4271.9 &  19(2985) &  13(19) \\
\hline
SPR &  100 &  CExt &  16 &  7 &  13(9) &  4(0) &  33515.3 &  4271.9 &  20(11853) &  13(16) \\
\hline
SPR &  100 &  RExt &  16 &  9 &  11(7) &  4(0) &  36026.8 &  4771.8 &  20(3067) &  15(20) \\
\hline
SPR &  100 &  RExt+CExt &  16 &  7 &  13(9) &  4(0) &  36026.8 &  4771.8 &  20(11617) &  13(16) \\
\hline
SPR &  200 &  No &  19 &  11 &  10(7) &  3(1) &  72890.0 &  10691.8 &  21(2755) &  16(23) \\
\hline
SPR &  200 &  CExt &  19 &  9 &  11(8) &  4(2) &  72890.0 &  10691.8 &  21(11094) &  15(21) \\
\hline
SPR &  200 &  RExt &  20 &  11 &  10(7) &  3(2) &  81023.0 &  13855.9 &  21(3951) &  16(23) \\
\hline
SPR &  200 &  RExt+CExt &  20 &  9 &  12(9) &  3(2) &  81023.0 &  13855.9 &  21(11119) &  15(21) \\
\hline
SPR &  300 &  No &  19 &  9 &  12(9) &  3(1) &  115421.1 &  14718.8 &  21(2994) &  15(22) \\
\hline
SPR &  300 &  CExt &  21 &  8 &  12(10) &  4(3) &  115421.1 &  18475.8 &  21(10483) &  14(20) \\
\hline
SPR &  300 &  RExt &  21 &  9 &  12(10) &  3(2) &  126820.0 &  20254.4 &  21(2254) &  16(22) \\
\hline
SPR &  300 &  RExt+CExt &  23 &  8 &  13(12) &  3(3) &  126820.0 &  23304.6 &  21(10729) &  15(20) \\
\hline
SPR &  2000 &  No &  20 &  5 &  11(9) &  8(6) &  851661.0 &  107984.2 &  16(809) &  10(12) \\
\hline
SPR &  2000 &  CExt &  22 &  4 &  11(10) &  9(8) &  851661.0 &  112221.5 &  16(6257) &  10(13) \\
\hline
SPR &  2000 &  RExt &  22 &  5 &  12(11) &  7(6) &  930431.6 &  131861.0 &  17(722) &  10(12) \\
\hline
SPR &  2000 &  RExt+CExt &  24 &  3 &  13(13) &  8(8) &  930431.6 &  133514.5 &  16(5921) &  10(13) \\
\hline
Prophet &  100 &  No &  16 &  13 &  6(2) &  5(1) &  33515.3 &  2096.9 &  19(2765) &  15(20) \\
\hline
Prophet &  100 &  CExt &  16 &  8 &  12(8) &  4(0) &  33515.3 &  2096.9 &  20(12357) &  14(15) \\
\hline
Prophet &  100 &  RExt &  16 &  13 &  7(3) &  4(0) &  36026.8 &  2705.5 &  20(2512) &  15(20) \\
\hline
Prophet &  100 &  RExt+CExt &  16 &  8 &  12(8) &  4(0) &  36026.8 &  2705.5 &  20(12029) &  14(15) \\
\hline
Prophet &  200 &  No &  19 &  15 &  7(4) &  2(0) &  72890.0 &  6004.4 &  22(2850) &  18(25) \\
\hline
Prophet &  200 &  CExt &  19 &  10 &  11(8) &  3(1) &  72890.0 &  6004.4 &  22(11345) &  17(21) \\
\hline
Prophet &  200 &  RExt &  20 &  15 &  7(4) &  2(1) &  81023.0 &  9781.2 &  22(2240) &  17(23) \\
\hline
Prophet &  200 &  RExt+CExt &  20 &  11 &  11(8) &  2(1) &  81023.0 &  9781.2 &  22(11303) &  17(20) \\
\hline
Prophet &  300 &  No &  19 &  14 &  8(5) &  2(0) &  115421.1 &  6744.1 &  22(2836) &  18(25) \\
\hline
Prophet &  300 &  CExt &  21 &  10 &  11(9) &  3(2) &  115421.1 &  8185.2 &  22(11303) &  17(21) \\
\hline
Prophet &  300 &  RExt &  21 &  13 &  9(7) &  2(1) &  126820.0 &  13161.6 &  22(1833) &  17(24) \\
\hline
Prophet &  300 &  RExt+CExt &  23 &  10 &  12(11) &  2(2) &  126820.0 &  14121.3 &  22(11243) &  17(21) \\
\hline
Prophet &  2000 &  No &  20 &  9 &  9(7) &  6(4) &  851661.0 &  79733.1 &  18(1347) &  12(17) \\
\hline
Prophet &  2000 &  CExt &  22 &  5 &  11(10) &  8(7) &  851661.0 &  76524.7 &  17(8291) &  11(14) \\
\hline
Prophet &  2000 &  RExt &  22 &  6 &  11(10) &  7(6) &  930431.6 &  84710.5 &  17(1312) &  11(16) \\
\hline
Prophet &  2000 &  RExt+CExt &  24 &  3 &  14(14) &  7(7) &  930431.6 &  81598.5 &  17(8030) &  11(14) \\
\hline
\end{tabular}

\caption{Patch Generation Results with Search Space Extensions}
\label{tab:results}
\end{table*}
}

Table~\ref{tab:results-nophp} presents a summary of the results for all of the
benchmarks except php. Table~\ref{tab:results-php} presents a summary of the
results for the php benchmark. Each row presents patch generation results for
SPR or Prophet with one search space configuration. The first column (System)
presents the evaluated system (SPR or Prophet). The second column (Loc. Limit)
presents the number of considered candidate program statements to patch under
the configuration. The third column (Space Extension) presents the
transformation  extensions that are enabled in the configuration: No (no
extensions, baseline search space), CExt (condition synthesis extension), RExt
(value replacement extension), or RExt+CExt (both).

The fourth column (Correct In Space) presents the number of defects with correct
patches that lie inside the full search space for the corresponding configuration. The
fifth column (Correct First) presents the number of defects for which the system
finds the correct patch as the {\em first} patch that validates against the 
test suite. 

Each entry of the sixth column (Plausible \& Blocked) is of the form X(Y).
Here X is the number of defects for which the system discovers a plausible but
incorrect patch as the first patch that validates. 
Y is the number of defects for which a plausible but incorrect patch 
blocked a subsequent correct patch (i.e., Y is the number of 
defects for which 1) the system discovers
a plausible but incorrect patch as the first patch that validates and 
2) the full search space contains a correct patch for that defect).

%the system also discovers a correct
%patch, and 3) the plausible but incorrect patch blocks the correct 
%patch (i.e., the system discovers the plausible but incorrect patch 
%before it discovers the correct patch). 

Each entry of the seventh column (Timeout) is 
also of the form X(Y). Here X is the number of defects for which the
system does not discover any plausible patch within the 12 hour timeout.
Y is the number of defects for which 1) the system does not discover a plausible
patch and 2) the full search space contains a correct patch for that defect. 

\comment{
Table~\ref{tab:results-nophp} presents results for the 11 non-php defects. 
Table~\ref{tab:results-php} presents results for the 13 php defects. The
fifth, sixth, and seventh columns in these two tables count each of these defects
into one of three bins: a correct patch as the first patch to 
validate (fifth column, Correct First), a plausible but incorrect
patch as the first patch to validate (X component, sixth column, 
Plausible \& Blocked), and no plausible patch found within the
12 hour timeout (X component, seventh
column, Timeout).  For configuration SPR 200 No in Table~\ref{tab:results-nophp},
for example, out of the 11 defects, SPR finds 2 correct patches as the first
to validate, 7 plausible but incorrect patches as the first to 
validate, and 2 timeouts with no plausible patch. 
}

The eighth column (Space Size) presents the
average number of candidate patch templates in the search space over all of
the 24 considered defects. 
Note that SPR and Prophet may instantiate multiple concrete patches with the
staged program repair technique from a patch template that contains an abstract
expression (See Section~\ref{sec:overview:overview}).
This column shows how the size of the search space grows
as a function of the number of candidate statements to patch and the
two extensions. Note that the CExt transformation extension does not 
increase the number of patch templates. Instead it increases the number
of concrete patches which each patch template generates.
The ninth column (Correct Rank) presents the
average rank of the first patch template that generates a correct patch in the search space over all of those
defects for which at least one correct patch is inside the search space.
Note that the correct rank increases as the size of the search space 
increases. 

Each entry of the tenth column (Plausible in 12h) is of the form X(Y).
Here X is the number of defects for which the system discovers a plausible
patch within the 12 hour timeout. Y is the sum, over the all of the
24 considered defects, of the number of plausible patches
that the system discovers within the 12 hour timeout. 

Each entry of the eleventh  column (Correct in 12h) is of the form X(Y).
Here X is the number of defects for which the system discovers a correct
patch (blocked or not) within the 12 hour timeout. Y is the number of correct 
patches that the system discovers within the 12 hour timeout. 

\subsection{Plausible and Correct Patch Density} 

An examination of the tenth column (Plausible in 12h) 
in Tables~\ref{tab:results-nophp} and \ref{tab:results-php}
highlights the overall plausible patch densities in the search spaces. 
For the benchmarks without php, the explored search spaces typically contain hundreds
up to a thousand plausible patches per defect. For php, in contrast, the explored
search spaces typically contain tens of plausible patches per defect. We 
attribute this significant difference in the plausible patch density to the 
quality of the php test suite and its resulting ability to successfully
filter out otherwise plausible but incorrect patches. Indeed, for three 
php defects, the php test suite is strong enough to filter out all of the
patches in the explored search spaces except the correct patch. 

An examination of the eleventh column (Correct in 12h) in Tables~\ref{tab:results-nophp} and ~\ref{tab:results-php}
highlights the overall correct patch densities in the explored search spaces. 
In sharp contrast to the plausible patch densities, the explored search spaces
contain, on average, less than two correct patches per defect for all 
of the benchmarks including php. There are five defects with as many as
two correct patches in any search space and one defect with as many as
four correct patches in any search space. The remaining defects contain
either zero or one correct patch across all of the search spaces. 

\subsection{Search Space Tradeoffs}

An examination of the fourth column (Correct In Space) in Table~\ref{tab:results-nophp} shows
that the number of correct patches in the full search space increases as the
size of the search space increases (across all benchmarks except php). 
But an examination of the fifth column (Correct First) indicates that 
that this increase does not translate into an increase in the ability
of SPR or Prophet to actually find these correct patches as the first
patch to validate. In fact, the ability of SPR and Prophet to isolate a
correct patch as the first patch to validate reaches a maximum at 
200 candidate statements with no extensions, then (in general) decreases
from there as the size of the search space increases. For php, Table~\ref{tab:results-php}
shows that the number of correct patches in the space does not 
significantly increase with the size of the search space, but that the
drop in the number of correct patches found as the first patch to 
validate is even more significant. Indeed, the 200+No Prophet configuration
finds 10 correct patches as the first patch to validate, while the 
largest 2000+RExt+CExt configuration finds only four! 

We attribute these facts to an inherent tradeoff in the search spaces. 
Expanding the search spaces to include more correct patches also
includes more implausible and plausible but incorrect patches.
The implausible patches consume validation time (extending the 
time required to find the correct patches), while the plausible
but incorrect patches block the correct patches. This trend is
visible in the Y entries in the sixth column in Table~\ref{tab:results-nophp} (Plausible \& Blocked) 
(these entries count the number of blocked correct patches), 
which generally increase as the size of the search space increases. 

Tables~\ref{tab:results-nophp} and \ref{tab:results-php} show how
this tradeoff makes the baseline SPR and Prophet configurations perform
best despite working with search spaces that contain fewer correct patches. 
Increasing the candidate statements beyond 200 never increases the number
of correct patches that are first to validate. Applying the 
CExt and RExt extensions also never increases the number of correct
patches that are first to validate. 

Our results highlight two challenges that SPR and Prophet 
(and other generate and validate systems) face when generating
correct patches:
\begin{itemize}\itemsep 0pt \parskip 0pt 
\sloppypar{
\item \textbf{Weak Test Suites: } The test suite provides incomplete
coverage.  The most obvious problem of the weak test suite is that it may accept incorrect
patches. Our results show that (especially for larger
search spaces) plausible but incorrect patches often block correct patches. 
For example, when we run Prophet with the baseline search space (200+No), there are only 4
defects whose correct patches are blocked; when we run Prophet with the largest
search space (2000+RExt+CExt), there are 11 defects whose correct patches are
blocked.
}

A more subtle problem is that weak test suites may increase the
validation cost of plausible but incorrect patches.
For such a patch, SPR or Prophet has to run the patched application on all
test cases in the test suite. If a stronger test suite is used, SPR and Prophet
may invalidate the patch with one test case and skip the remaining test cases. 

\sloppypar{
\item \textbf{Search Space Explosion: } A large search space contains many
candidate patch templates and our results show that it may be intractable to
validate all of the candidates. For example, with the baseline search space (200+No),
Prophet times out for only two defects (whose correct patches are outside the
search space); with the largest evaluated search space (2000+RExt+CExt), Prophet times out for
seven defects (whose correct patches are inside the search space).
}
\end{itemize}

Note that many previous systems~\cite{LeGoues:2012, Weimer:2013, Kim:2013,
Qi:2014} neglect the weak test suite problem and do not evaluate whether the
generated patches are correct or not. In contrast, our results show that the
weak test suite problem is at least as important as the search space explosion
problem. In fact, for all evaluated search space configurations, there are more
defects for which SPR or Prophet generates plausible but incorrect patches than
for which SPR or Prophet times out. 

\comment{
\subsection{Correct and Plausible Patches}

Our results highlight the difference of the distributions of plausible
patches and correct patches in the search space. The correct patches for
the defects in our benchmark set are sparse. For 45 of the 69 defects, all of
the considered search spaces contains no correct patch. For 15 out of the
remaining 24 defects, there is at most one correct patch in each of the search
spaces. For each of the remaining 9 defects, there are at most four correct
patches in each of the search spaces.

As the search space expands, the number of candidate patch templates increases
from the baseline size XXXX to XXXXX. In contrast, the number of correct patches
that SPR and Prophet generates within 12 hours slightly decreases from 23 and
25 to 13 and 15, respectively. Therefore one possible explanation of why larger
search spaces do not enable SPR or Prophet to generate more correct patch is
that the larger search spaces tend to have {\em lower correct patch density}.

Comparing to the correct patches, the plausible patches are relatively
abundant. Even with the baseline search space, SPR and Prophet is able to find
XXXX and XXXX plausible patches for the 24 defects within 12 hours,
respectively. There are XX of the 24 defects for which the search spaces may
contain more than YY plausible patches. 

Moreover, as the search space extends, the number of generated plausible
patches within 12 hours often increases, especially if we consider more
sophisticated concrete conditions (CExt). This is because with a weak test
suite, complicated transformations tend to generate more plausible but
incorrect patches similar to an existing plausible patch. These plausible
patches are pseudo-equivalent as the test suite cannot distinguish them.
Therefore another possible explanation of why larger search spaces do not
enable SPR and Prophet to generate correct patches for more defects is that
{\em more pseudo-equivalent plausible patches} make identifying correct patches
harder.
}

\subsection{SPR and Prophet Effectiveness}

\if 0
\begin{table}
\small
\input{figs/tab-comp-nophp-3.tex}
\caption{Correct Patch Selection Probability (excluding php), top 3}
\label{tab:comparison-nophp}
\vspace{-2mm}
\end{table}

\begin{table}[!t]
\small
\input{figs/tab-comp-php-3.tex}
\caption{Correct Patch Selection Probability (php only), top 3}
\label{tab:comparison-php}
\vspace{-2mm}
\end{table}

\begin{table}
\small
\input{figs/tab-comp-nophp-5.tex}
\caption{Correct Patch Selection Probability (excluding php), top 5}
\label{tab:comparison-nophp}
\vspace{-2mm}
\end{table}

\begin{table}[!t]
\small
\input{figs/tab-comp-php-5.tex}
\caption{Correct Patch Selection Probability (php only), top 5}
\label{tab:comparison-php}
\vspace{-2mm}
\end{table}
\fi

\begin{table}
\small
\begin{tabular}{|l|c|c|c|c|}
\hline
\multirow{2}{*}{\textbf{Search Space}} & \multirow{2}{*}{\scriptsize \textbf{SPR}} & \multirow{2}{*}{\scriptsize \textbf{Prophet}} & \textbf{\scriptsize Random} & \textbf{\scriptsize Random} \\
& & & \textbf{\scriptsize (SPR)} & \textbf{\scriptsize (Prophet)} \\
\hline
{\tiny 100+No} & 52 / 3 & 38 / 4 & 65.0 / 1.4 & 65.0 / 1.4 \\
\hline
{\tiny 100+CExt} & 65 / 2 & 53 / 3 & 73.0 / 1.0 & 73.0 / 1.0 \\
\hline
{\tiny 100+RExt} & 52 / 3 & 38 / 4 & 65.1 / 1.4 & 65.1 / 1.4 \\
\hline
{\tiny 100+RExt+CExt} & 65 / 2 & 53 / 3 & 73.0 / 1.0 & 73.0 / 1.0 \\
\hline
{\tiny 200+No} & 59 / 4 & 45 / 5 & 73.5 / 2.7 & 76.2 / 2.1 \\
\hline
{\tiny 200+CExt} & 66 / 3 & 54 / 4 & 78.1 / 1.7 & 78.1 / 1.7 \\
\hline
{\tiny 200+RExt} & 59 / 4 & 45 / 5 & 76.2 / 2.1 & 75.9 / 2.1 \\
\hline
{\tiny 200+RExt+CExt} & 66 / 3 & 54 / 4 & 77.8 / 1.7 & 77.8 / 1.7 \\
\hline
{\tiny 300+No} & 60 / 5 & 50 / 5 & 80.2 / 2.0 & 78.2 / 2.1 \\
\hline
{\tiny 300+CExt} & 75 / 2 & 63 / 3 & 83.9 / 1.3 & 83.2 / 1.4 \\
\hline
{\tiny 300+RExt} & 62 / 4 & 50 / 5 & 81.4 / 1.4 & 79.9 / 2.1 \\
\hline
{\tiny 300+RExt+CExt} & 75 / 2 & 63 / 3 & 85.2 / 1.1 & 84.4 / 1.2 \\
\hline
{\tiny 2000+No} & 56 / 4 & 50 / 5 & 78.9 / 1.3 & 77.8 / 2.3 \\
\hline
{\tiny 2000+CExt} & 72 / 2 & 63 / 3 & 86.6 / 0.7 & 83.2 / 1.4 \\
\hline
{\tiny 2000+RExt} & 60 / 3 & 51 / 5 & 74.7 / 1.7 & 78.5 / 2.1 \\
\hline
{\tiny 2000+RExt+CExt} & 72 / 2 & 64 / 3 & 82.6 / 1.1 & 84.4 / 1.3 \\
\hline
\end{tabular}

\caption{Costs and Payoffs of Reviewing the First 10 Generated Patches (excluding php)}
\label{tab:comparison-nophp}
\vspace{-2mm}
\end{table}

\begin{table}[!t]
\small
\begin{tabular}{|l|c|c|c|c|}
\hline
\multirow{2}{*}{\textbf{Search Space}} & \multirow{2}{*}{\scriptsize \textbf{SPR}} & \multirow{2}{*}{\scriptsize \textbf{Prophet}} & \textbf{\scriptsize Random} & \textbf{\scriptsize Random} \\
& & & \textbf{\scriptsize (SPR)} & \textbf{\scriptsize (Prophet)} \\
\hline
{\tiny 100+No} & 38 / 9 & 37 / 9 & 45.8 / 8.1 & 44.7 / 8.4 \\
\hline
{\tiny 100+CExt} & 48 / 9 & 56 / 9 & 66.7 / 6.8 & 66.8 / 6.8 \\
\hline
{\tiny 100+RExt} & 39 / 10 & 39 / 10 & 50.9 / 8.8 & 51.4 / 8.9 \\
\hline
{\tiny 100+RExt+CExt} & 46 / 9 & 57 / 9 & 66.9 / 6.8 & 66.7 / 6.8 \\
\hline
{\tiny 200+No} & 39 / 10 & 39 / 11 & 47.7 / 9.1 & 49.3 / 10.4 \\
\hline
{\tiny 200+CExt} & 39 / 10 & 57 / 11 & 59.7 / 7.6 & 68.1 / 7.9 \\
\hline
{\tiny 200+RExt} & 39 / 10 & 40 / 11 & 47.8 / 9.1 & 51.6 / 10.1 \\
\hline
{\tiny 200+RExt+CExt} & 39 / 10 & 58 / 11 & 59.6 / 7.6 & 68.0 / 7.9 \\
\hline
{\tiny 300+No} & 32 / 9 & 39 / 11 & 43.8 / 8.1 & 50.2 / 10.4 \\
\hline
{\tiny 300+CExt} & 32 / 9 & 57 / 11 & 59.3 / 6.4 & 72.8 / 7.9 \\
\hline
{\tiny 300+RExt} & 34 / 9 & 40 / 11 & 45.8 / 8.1 & 51.5 / 10.2 \\
\hline
{\tiny 300+RExt+CExt} & 34 / 9 & 58 / 11 & 61.3 / 6.4 & 69.5 / 8.0 \\
\hline
{\tiny 2000+No} & 7 / 5 & 25 / 7 & 16.7 / 4.4 & 37.4 / 6.3 \\
\hline
{\tiny 2000+CExt} & 7 / 5 & 34 / 5 & 29.8 / 2.9 & 46.4 / 4.0 \\
\hline
{\tiny 2000+RExt} & 9 / 5 & 17 / 6 & 18.7 / 4.4 & 29.3 / 5.3 \\
\hline
{\tiny 2000+RExt+CExt} & 7 / 5 & 27 / 5 & 29.8 / 2.9 & 47.4 / 3.2 \\
\hline
\end{tabular}

\caption{Costs and Payoffs of Reviewing the First 10 Generated Patches (php only)}
\label{tab:comparison-php}
\vspace{-2mm}
\end{table}

We compare the effectiveness of the SPR and Prophet patch prioritization orders
by measuring the costs and payoffs for a human developer who reviews the generated
patches to find a correct patch.  We consider a scenario in which the developer reviews the first 10 generated patches one by one for each defect until he finds a correct patch.
He gives up if none of the first 10 patches are correct. 
For each system
and each search space configuration, we compute
(over the 24 defects that have correct patches in the full SPR and Prophet search space)
1) the total number of patches
the developer reviews (this number is the cost) and 2) the total number
of defects for which the developer obtains a correct patch (this number is the payoff). 
We also compute the expected costs and payoffs if the developer examines the generated
plausible SPR and Prophet patches in a random order. 
The raw data used to compute these numbers is available at Appendix.

Tables~\ref{tab:comparison-nophp} and \ref{tab:comparison-php} present
these costs and payoffs.  The first column
presents the search space configuration. The second and third columns present
the costs and payoffs for the SPR and Prophet patch prioritization orders;
the fourth and fifth columns present the corresponding costs and payoffs
for the random orders. 
Each entry is of the form X/Y, where X is the total
number of patches that the developer reviews and Y is
the total number of defects for which he obtains a correct patch.
These numbers highlight the effectiveness of the SPR and Prophet patch
prioritization in identifying correct patches within much larger
sets of plausible but incorrect patches. 

\comment{
Both SPR and Prophet prioritization order
significantly outperforms a random order. Moreover, Prophet prioritization
order enables the developer to obtain correct patches for more defects via
reviewing less patches than SPR.
}

\if 0
{\em correct patch selection probability} ---
i.e., the probability of getting a correct patch when
1) first randomly selecting a defect with at least one plausible
patch in the explored space then 2) choosing the first
patch to validate for that defect. Our random comparison is
the probability of getting a correct patch when the second
choice (the first patch to validate) is replaced with a random
selection of a plausible patch from the set of plausible patches 
in the explored space. The raw data used to compute these
numbers is available at \url{http://groups.csail.mit.edu/pac/ss/}.

Table~\ref{tab:comparison-nophp} presents the
correct patch selection probabilities for the benchmarks
without php. These numbers highlight the effectiveness of
the SPR and Prophet patch prioritization mechanisms 
at isolating correct patches within a sea of plausible
but incorrect patches, with Prophet roughly twice as
effective as SPR at isolating a correct patch. For the
baseline 200+No configuration, Prophet finds the correct
patch more than 50\% of the time, SPR more than 20\%
of the time, while random finds a correct patch less
than 10\% of the time. The random comparison for the
200 configurations is significantly higher than for
other configurations because, for 
one defect (gmp-13420-13421), all or all but one
of the plausible patches are correct. For other 
configurations there are either no plausible patches
(so the defect is not included)
or there are many more plausible but correct patches. 
\fi

Table~\ref{tab:comparison-php} presents the corresponding results for the php
defects. These numbers highlight the difference that a stronger test suite can
make in the success of finding correct patches. The correct patch selection
probabilities are dramatically higher for php than for the other benchmarks.
But note that as the patch search spaces become large, the number of defects
for which the developer obtains correct patches become smaller, reflecting 1)
the increasing inability of the systems to find any correct patch in the
explored space within the 12 hour timeout and 2) the increasing presence of
blocking plausible but incorrect patches. 

Finally, these numbers highlight the effectiveness of the Prophet learned patch
prioritization across the board --- the developer always obtains correct
patches for at least as many defects with Prophet as with SPR.

%Prophet exhibits a larger correct
%patch selection probability than SPR. And Prophet always (except for the
%largest 2000+RExt+CExt configuration) finds more correct patches as the first
%patch to validate. 

\comment{
To determine the patch prioritization order, SPR uses a set of hand-coded
heuristics (see Section~\ref{XXX} and \cite{SPR}), while Prophet learns a
probabilistic patch correctness model from a large database of previous
successful human patches (see Section~\ref{XXX} and \cite{ProphetTR,Prophet}).
}

\comment{
\begin{table}
\small
\begin{tabular}{|l|c|c|c|c|}
\hline
\multirow{2}{*}{\textbf{Search Space}} & \multirow{2}{*}{\textbf{SPR}} & \multirow{2}{*}{\textbf{Prophet}} & \textbf{Random} & \textbf{Random} \\
& & & \textbf{(SPR)} & \textbf{(Prophet)} \\
\hline
100+No & 47.4\% & 63.2\% & 31.7\% & 31.9\% \\
\hline
100+CExt & 35.0\% & 40.0\% & 26.9\% & 26.9\% \\
\hline
100+RExt & 45.0\% & 65.0\% & 30.9\% & 30.6\% \\
\hline
100+RExt+CExt & 35.0\% & 40.0\% & 26.9\% & 26.9\% \\
\hline
200+No & 52.4\% & 68.2\% & 33.8\% & 33.1\% \\
\hline
200+CExt & 42.9\% & 45.5\% & 30.6\% & 29.4\% \\
\hline
200+RExt & 52.4\% & 63.6\% & 35.1\% & 34.1\% \\
\hline
200+RExt+CExt & 42.9\% & 45.5\% & 32.2\% & 30.9\% \\
\hline
300+No & 42.9\% & 63.6\% & 26.9\% & 27.9\% \\
\hline
300+CExt & 38.1\% & 45.5\% & 22.9\% & 22.7\% \\
\hline
300+RExt & 42.9\% & 59.1\% & 26.5\% & 27.7\% \\
\hline
300+RExt+CExt & 38.1\% & 45.5\% & 22.8\% & 23.3\% \\
\hline
2000+No & 31.2\% & 50.0\% & 18.0\% & 24.9\% \\
\hline
2000+CExt & 25.0\% & 29.4\% & 13.8\% & 19.5\% \\
\hline
2000+RExt & 29.4\% & 41.2\% & 17.7\% & 20.8\% \\
\hline
2000+RExt+CExt & 18.8\% & 23.5\% & 14.6\% & 13.9\% \\
\hline
\end{tabular}

\caption{Patch Prioritization Comparison}
\label{tab:comparison}
\end{table}
}

\comment{
Table~\ref{tab:comparison} presents the comparison of different patch
prioritization orders for the 24 tested defects under each search space
configuration. The first column (Search Space) presents the search space
configurations. The integer number in the each entry corresponds to the number
of considered program statements returned by the error localizer. The presences
of ``CExt'' and ``RExt'' indicate whether the corresponding search space
extensions are enabled or not. The second column (Prophet) and the third column
(SPR) present the probabilities for which Prophet and SPR generate a correct
patch as the first patch, respectively. The fourth column (Random (Prophet))
and the fifth column (Random (SPR)) present the corresponding probabilities, as
if Prophet and SPR use a random prioritization order to sort the generated
plausible patches instead, respectively.

Our results show that the prioritization orders of SPR and Prophet are
effective for the defects in our benchmark set. The orders of SPR and Prophet
are up to XX times more likely to isolate the correct patches among many
plausible patches than a random prioritization order.

The results also show that the prioritization order of Prophet outperforms both
random and SPR under all search space configurations. This indicates that the
learned probabilistic model is useful and enables Prophet to prioritizes
correct patches before all plausible but incorrect patches for more cases.
}

\comment {
We compare the repair generation results of various search space 
configurations of SPR and Prophet to answer the RQ1:
\begin{quote}
\framebox{
\parbox{0.39\textwidth}{
\noindent{\bf RQ1:}
Will exploring larger search space enable SPR and Prophet to generate correct
patches for more defects?
}
}
\end{quote}

\noindent \textbf{Answer to RQ1: } 
The results show that a larger search space will always includes more correct
repairs into the search space, but not always enables the system to
automatically generate more correct repairs. In fact, SPR and Prophet generate
correct repairs for the maximum number of defects (11 and 15 defects,
respectively) with the baseline search space configuration.
The results show that exploring a larger space causes SPR and Prophet
generate correct repairs for less defects in our benchmark set.

\begin{quote}
\framebox{
\parbox{0.39\textwidth}{
\noindent{\bf RQ1a:}
Will considering more candidate program statements from the error localization 
results enable SPR and Prophet to generate correct patches for more defects?
}
}
\end{quote}

\noindent \textbf{Answer to RQ1a: } 
Our results show an inherent trade-off of selecting the number of candidate
program statements to modify. On one hand, if we only select the first 100
program statements to modify (instead of the first 200), correct repairs for
three defects will be excluded from the resulting search space. Therefore SPR
and Prophet will generate only XX and 12 defects (instead of 11 and 15
defects), respectively. On the other hand, if we select more statements (i.e., 300 or 2000),
the resulting search space will be too large so that SPR and Prophet generate
correct repairs for less defects.

\begin{quote}
\framebox{
\parbox{0.39\textwidth}{
\noindent{\bf RQ1b:}
Will considering more mutation operations enable SPR and Prophet to
automatically generate correct patches for more defects?
}
}
\end{quote}

\noindent \textbf{Answer to RQ1b: } 

\subsection{Key Challenges}

\begin{quote}
\framebox{
\parbox{0.39\textwidth}{
\noindent{\bf RQ2:}
What are the key challenges for SPR and Prophet to automatically identify and generate
correct patches in the enlarged search space?
}
}
\end{quote}

If a SPR or Prophet run does not automatically generate a correct repair for a
defect, then it either generates a plausible but incorrect repair or fails to
generate any repair that passes the supplied test suite within 12 hours. 

\noindent \textbf{Answer to RQ2:} Our results identify two key challenges for
SPR and Prophet to generate correct repairs:
\begin{itemize}
\item \textbf{Weak Test Suite: } The user supplied test suite is incomplete.
The most obvious problem of the weak test suite is that it may accept incorrect
repairs. Our results show that SPR or Prophet often finds a plausible but
incorrect repair that passes the test suite before it reaches a correct repair,
i.e. The correct repair is blocked by the plausible but incorrect repair. Our
results further show that as the search space gets larger, the correct repairs
for more defects are tend to be blocked by plausible but incorrect repairs. For
example, when we run Prophet with the baseline search space, there are only 4
defects whose correct repairs are blocked; when we run Prophet with the largest
search space in our experiments, there are 14 defects whose correct repairs are
blocked.

A more subtle problem of the weak test suite is that it may increase the
validation cost of SPR or Prophet for those plausible but incorrect repairs.
For such a repair, SPR or Prophet has to run the repaired application on all
test cases in the test suite. If a stronger test suite is used, SPR and Prophet
may invalidate the repair with one test case and skip the remaining test cases. 

\item \textbf{Search Space Explosion: } A large search space contains many
candidate repair templates and our results show that it may be intractable to
validate all of the candidates. For example, with the baseline search space,
Prophet times out for only two defects (whose correct patches are outside the
search space); with the largest search space, Prophet times out for seven
defects (whose correct patches are even inside the search space).

%cannot find any plausible repair that passes the test suite within 12
%hours for two defects (whose correct repairs are outside the search space);
%with the largest evaluated search space, Prophet cannot find any plausible
%repair for seven defects (whose correct repairs are inside the search space). 

\end{itemize}

Note that many previous systems~\cite{LeGoues:2012, Weimer:2012, Kim:2013,
Qi:2014} neglect the weak test suite problem and do not evaluate whether the
generated patches are correct or not. In contrast, our results show that the
weak test suite problem is at least as important as the search space explosion
problem. In fact, for all evaluated search space configurations, there are more
benchmark defects for which SPR or Prophet generates plausible but incorrect
repairs than for which SPR or Prophet times out.

\subsection{Results Explanation}

\begin{quote}
\framebox{
\parbox{0.39\textwidth}{
\noindent{\bf RQ3:}
Why a larger search space does not enable SPR or Prophet to generate 
correct repairs for more defects?
}
}
\end{quote}

To answer the above question, we summarize important statistics of each of the
repair generation runs under different search space configurations. 

\noindent \textbf{Answer to RQ3: } A larger search space often does not enable
SPR or Prophet correct repairs for more defects because of the following two 
reasons:
\begin{itemize}
\item \textbf{Lower Correct Repair Density: } 

\item \textbf{More Pseudo-equivalent Repairs: }
\end{itemize}

Lower density of correct repairs. Higher
ratio of plausible/correct patches. Validation order / additional information
is important.

RQ3a

\noindent \textbf{Answer to RQ3a: }

RQ3b

\noindent \textbf{Answer to RQ3b: }

\subsection{Prophet}

RQ4

\noindent \textbf{Answer to RQ4: } Validation order is important. Identify the correct 
repairs among multiple plausible repairs.
}

\section{Threats to Validity}
\label{sec:threat}

This paper presents a systematic study of search space tradeoffs with SPR and
Prophet. One threat to validity is that our results will not generalize to
other benchmark sets and other patch generation systems. Note that the
benchmark set was developed by other researchers, not by us, with the
goal of obtaining a large, unbiased, and realistic benchmark set~\cite{LeGoues:2012}.
And this same benchmark set has been used to evaluate many previous patch generation
systems~\cite{LeGoues:2012, Weimer:2013, Qi:2014, Kali, SPR}. 
The observations in this paper are consistent with previous results
reported for other systems on this benchmark set~\cite{Kali,LeGoues:2012,Weimer:2013,Qi:2014}.

Another threat to validity is that stronger test suites will become
the norm so that the results for benchmark applications other than php will not
generalize to other applications. We note that 1) comprehensive test coverage is
widely considered to be beyond reach for realistic applications, and 2) 
even php, which has by far the strongest test suite in the
set of benchmark applications, has multiple defects for which the number of plausible patches
exceeds the number of correct patches by one to two orders of magnitude. 

\if 0
\input{paper_related.tex}
\fi
\section{Related Work}
\label{sec:related_work}

\noindent{\bf ClearView:} 
ClearView is a generate-and-validate system that observes normal executions to
learn invariants that characterize safe behavior~\cite{Perkins:2009}. It deploys
monitors that detect crashes, illegal control transfers and out of bounds write
defects. In response, it selects a nearby invariant that the input that
triggered the defect violates, and generates patches that take a repair action
to enforce the invariant. 

A Red Team evaluation found that ClearView was able to automatically generate patches
that eliminate 9 of 10 targeted Firefox vulnerabilities and 
enable Firefox to continue to execute successfully~\cite{Perkins:2009}. 
For 5 of these 9 defects, the first patch that ClearView generated
was successful. For 1 of the 9, the second patch was successful.
For 2 of the 9, the third patch was successful. The
final defect that ClearView was able to patch required 3 distinct
ClearView patches --- each of the first 2 ClearView patches exposed
a new defect that ClearView then patched. For these 3 defects, the
first patch was successful. These numbers illustrate the density
with which successful patches appear in the ClearView search space. 
We attribute this density, in part, to the fact that ClearView leverages the learned invariant
information to focus the search on successful patches and does not rely 
solely on the validation test suite. 

\noindent{\bf Kali:} Kali is a generate-and-validate system that 
deploys a very simple strategy --- it simply removes functionality. This strategy is obviously not
intended to correctly repair a reasonable subset of the defects that occur
in practice. Nevertheless, the results show that this strategy is at least
as effective as previous more involved repair strategies that aspire to successfully
repair a broad class of defects~\cite{Kali}. The standard scenario is that
the defect occurs in a region of code that one of the negative test cases
but few if any of the positive test cases exercises. Kali simply deletes
the region of code to produce a program that passes the validation test
suite but immediately fails on new test cases that exercise the removed
functionality. 

These results are consistent with the results presented in this paper, highlight the
inadequacy of current test suites to successfully identify incorrect patches,
and identify one prominent source of the relatively many plausible but incorrect patches 
that occur in current patch search spaces. 

A recent paper~\cite{NOPOLTR} evaluates GenProg patches,
NOPOL~\cite{DeMarco:2014} patches, and Kali remove statement patches for 224
Java program defects in the Defects4J dataset~\cite{Defects4J}. The results
are, in general, consistent with our results in this paper. 
Out of 42 manually analyzed plausible patches, the analysis indicates that only 8
patches are undoubtedly correct.

%This paper presents, to the best of our knowledge, the first systematic study
%of the search space tradeoffs in automatic patch generation systems. There are
%many patch generation and program repair techniques and there are also a plenty
%of previous studies on other aspects of existing patch generation systems. 

\noindent{\bf GenProg, AE, and RSRepair:} GenProg~\cite{LeGoues:2012}, AE~\cite{Weimer:2013}, 
and RSRepair~\cite{Qi:2014} were all evaluated on (for RSRepair, a subset of)
the same benchmark set that we use
in this paper to evaluate the SPR and Prophet search spaces. The presented results
in the GenProg, AE, and RSRepair papers~\cite{LeGoues:2012, Weimer:2013, Qi:2014} 
focus on the ability of the systems to generate plausible patches for the defects
in the benchmark set. There is no attempt to determine whether the generated patches
are correct or not.  Unfortunately, the presented evaluations of these systems suffer 
from the fact that the testing
infrastructure used to validate the candidate patches contains errors that
cause the systems to incorrectly accept implausible patches that do not even pass all of the test
cases in the validation test suite~\cite{Kali}. 

\sloppypar{
A subsequent study of these systems corrects these errors and sheds more light
on the subject~\cite{Kali}. This study found that 1) the systems generate
correct patches for only 2 (GenProg, RSRepair) or 3 (AE) of the 105 benchmark
defects/functionality changes in this benchmark set, 2) the systems generate
plausible but incorrect patches for 16 (GenProg), 8 (RSRepair), and 24 (AE)
defects/functionality changes, and 3) the majority of the plausible patches,
including all correct patches, are equivalent to a single modification that
deletes functionality. Moreover, the correct SPR and Prophet patches for these
defects lie outside the GenProg, AE, and RSRepair search space (suggesting that
these systems will never be able to generate a correct patch for these
defects)~\cite{SPR}.
}

These results are broadly consistent with the results presented in this paper.
The study found that only 5 of the 110 plausible GenProg patches are correct,
only 4 of the analyzed plausible 44 RSRepair patches are correct, and only 3 of the 27
plausible AE patches are correct~\cite{Kali}. These results are consistent with
the hypothesis that, in the GenProg, AE, and RSRepair search space, as in the
SPR and Prophet search spaces, plausible but incorrect patches occur more
abundantly than correct patches. A reasonable conclusion is that the GenProg,
AE, and RSRepair search space, while containing fewer correct
and plausible patches than the richer SPR and Prophet search
spaces~\cite{SPR,Prophet}, still exhibits the basic pattern of sparse correct
patches and more abundant plausible but incorrect patches. 

A subsequent study of GenProg and RSRepair (under the name TrpAutoRepair) on small student programs
provides further support for this hypothesis~\cite{Overfitting}. The results indicate that patches
validated on one test suite typically fail to generalize to produce correct results
on other test suites not used to validate the patches. These results are consistent with 
a GenProg/RSRepair search space that contains sparse correct patches and more
abundant plausible but incorrect patches. 

We note that RSRepair uses random search; previous
research found that the GenProg genetic search algorithm 
performs no better than random search on a subset of the benchmarks~\cite{Qi:2014}. All of the published research
is consistent with the hypothesis that, because of the relative abundance
of plausible but incorrect patches, systems (such as GenProg and RSRepair)
that perform no better than random search will need to incorporate 
additional sources of information other than the validation test suite
if they are to successfully generate correct patches in the presence
of current relatively weak test suites.\footnote{
Of course, if the patch space does not contain correct patches,
stronger test suites will prevent the system from generating
any patches at all. This is the case for GenProg ---
at most two additional test cases per defect completely disable GenProg's
ability to produce any patch at all except the five correct patches
in its search space~\cite{Kali}.
}

\noindent{\bf NOPOL:}
NOPOL~\cite{DeMarco:2014, NOPOLTR} is an automatic repair tool focusing on branch
conditions. It identifies branch statement directions that can pass negative
test cases and then uses SMT solvers to generate repairs for the branch
condition. 

\noindent{\bf Data Structure Repair:} Data structure repair enables applications
to recover from data structure corruption errors~\cite{DemskyR03,DemskyR06}. It
enforces a data structure consistency specification. This specification
can be provided by a developer or automatically inferred from correct
program executions~\cite{DemskyEGMPR06}.

In general, there are multiple repaired data structures that satisfy the
consistency specification.  Data structure repair operates over search space 
generated by a set of repair operations to generate a repaired data structure
that is heuristically close to the original unrepaired data structure. The
goal is not to obtain a hypothetical data structure, but instead a consistent data
structure that enables acceptable continued execution. 

\noindent {\bf PAR:}
PAR~\cite{Kim:2013} is another prominent automatic patch generation system. PAR
is based on a set of predefined human-provided patch templates. 
We are unable to directly compare PAR with other
systems because, despite repeated requests to the authors of the PAR paper over
the course of 11 months, the authors never provided us with the patches that
PAR was reported to have generated~\cite{Kim:2013} or the necessary materials to
would have enabled us to reproduce the PAR experiments.

\sloppypar{
The PAR search space
(with the eight templates in the PAR paper~\cite{Kim:2013}) is in fact a subset
of the SPR and Prophet search space. Monperrus found that PAR fixes the
majority of its benchmark defects with only two templates (``Null Pointer
Checker'' and ``Condition Expression
Adder/Remover/Replacer'')~\cite{Monperrus:2014}. 
}

%Monperrus found that PAR fixes the majority of its
%benchmark defects with only two templates (``Null Pointer Checker'' and
%``Condition Expression Adder/Remover/Replacer'')~\cite{Monperrus:2014}. 

\noindent{\bf Plastic Surgery Hypothesis: }
Barr et al.~\cite{Barr:2014} studied the platic surgery hypothesis, i.e.,
changes to a codebase contain snippets that already exist in the codebase at
the time of the changes. The effectiveness of copy mutations of many patch
generation systems relies on this hypothesis. Their results show that 10\% of the
commits can be completely constructed from existing code.  

\noindent {\bf Fix Ingredient Availability: }
Martinez et. al.~\cite{Martinez:2014} studied more than 7,000 human commits in
six open source programs to measure the fix ingredient availability,
i.e., the percentage of commits that could be synthesized solely from existing
lines of code.  The results show that 3-17\% of the commits can be
synthesized from existing lines of code. 
%The study provides another potential
%explanation for the inability of GenProg, RSRepair, and AE to generate correct
%patches --- the GenProg, RSRepair and AE search space only contains patches
%that can be synthesized from existing lines of code (specifically by copying
%and/or removing existing statements without variable replacement or any other
%expression-level modification).

%\noindent{\bf Maintainability: }
%A paper investigating the developer maintainability of a subset of the 
%GenProg patches found ``statistically significant evidence'' that
%the GenProg patches ``can be maintained with equal accuracy and less effort
%than the code produced by human-written patches''~\cite{fry2012human}. 
%In retrospect, this potentially surprising result may become more plausible
%when one considers that the machine-generated patches 
%are equivalent to a single functionality deletion modification. 
%XXX However, they are evaluating incorrect patches.

\noindent{\bf Real World Fix: } Zhong and Su systematically analyzed more than
9000 real world bug fixes and studied several research questions including how
many locations each bug fix modifies and what modification operators are
essential for bug fixing~\cite{Zhong:2015}. 

\noindent{\bf Repair Model: } Martinez and Monperrus manually mined previous
human patches and suggest that if a patch generation system works with a
non-uniform probabilistic model, the system would find plausible patches
faster~\cite{MiningRepairModel}. In contrast, Prophet~\cite{Prophet}
automatically learns a probabilistic model from past successful patches.
Prophet is the first patch generation system to operate with such a learned
model~\cite{Prophet} to identify potentially correct patches.

\noindent{\bf SemFix and MintHint:}
SemFix~\cite{Nguyen:2013} and MintHint~\cite{Kaleeswaran:2014} replace the
faulty expression with a symbolic value and use symbolic execution
techniques~\cite{KLEE} to find a replacement expression that
enables the program to pass all test cases. SemFix and MintHint are 
evaluated only on applications with less than 10000 lines of code. 
In addition, these techniques cannot generate fixes for statements
with side effects. 

\noindent \textbf{CodePhage:}
Horizontal code transfer automatically locates correct code in one application,
then transfers that code into another application~\cite{Sidiroglou-Douskos15}.
This technique has been applied to eliminate otherwise fatal integer overflow,
buffer overflow, and divide by zero errors and shows enormous potential for
leveraging the combined talents and labor of software development efforts
worldwide. 

\noindent{\bf Defect Repair via Q\&A Sites: } Gao et. al.~\cite{QingGaoASE15}
propose to repair recurring defects by analyzing Q\&A sites such as Stack
Overflow. The proposed technique locates the relevant Q\&A page for a recurring
defect via a search engine query, extracts code snippets from the page, and
renames variables in the extracted code snippets to generate patches. This
technique addresses the weak test suite problem with the additional information
collected from the Q\&A sites, but it requires the existence of exact defect
repair logic on Q\&A pages.

%{\name} is
%different from this technique, because {\name} directly learns a probabilistic
%model from successful human patches and does not rely on the existence of exact
%defect repair logic on Q\&A pages.

%\noindent{\bf Debroy and Wong:}
%Debroy and Wong~\cite{Debroy:2010} present a transformation-based patch generation
%technique. This technique either replaces an arithmetic operator with another
%operator or negates a condition. In contrast, {\name} uses more sophisticated
%and effective transformations and search algorithms. None of the correct
%repairs in {\name}'s search space for the 19 defects are within the Debroy and
%Wong search space. 

\noindent{\bf AutoFixE: }
AutoFix-E~\cite{AutoFix-E, AutoFixETSE} operates with a set of fix schemas to repair Eiffel
programs with human-supplied specifications called contracts. 

%\noindent{\bf Deductive Program Repair: }
%Deductive Program Repair formalizes the program repair problem as a program
%synthesis problem, using the original defective program as a
%hint~\cite{DeductiveProgramRepair}. It replaces the expression to repair with a
%synthesis hole and uses a counterexample-driven synthesis algorithm to find a
%patch that satisfies a formal specification. 

\sloppypar{
\noindent{\bf Domain Specific Repair Generation: }
Other program repair systems include VEJOVIS~\cite{Ocariza:2014} and Gopinath
et al.~\cite{Gopinath:2014}, which applies domain specific techniques to repair
DOM-related faults in JavaScript and selection statements in database programs
respectively. 
}

\noindent{\bf Failure-Oblivious Computing:}
Failure-oblivious computing~\cite{Rinard:2004} checks for out of bounds reads and
writes. It discards out of bounds writes and manufactures values for out
of bounds reads. This eliminates data corruption from out of bounds writes,
eliminates crashes from out of bounds accesses, and enables the program
to continue execution along its normal execution path.

\noindent{\bf RCV:}
RCV~\cite{Long:2014} enables applications to survive null dereference and divide by zero errors.
It discards writes via null references, returns zero for reads via null references,
and returns zero as the result of divides by zero. Execution continues along the
normal execution path.

\noindent{\bf Bolt:}
Bolt~\cite{Kling:2012} attaches to a running application, determines if the application
is in an infinite loop, and, if so, exits the loop. A user can also use Bolt to
exit a long-running loop.  In both cases the goal is to enable the application
to continue useful execution. Bolt uses checkpoint and restore to explore a 
search space of loop exit strategies, including jumping to the instruction
immediately following the loop as well as returning to one of the (in
general many) invoked procedures on the call stack. 

\noindent{\bf Cyclic Memory Allocation:} Cyclic memory allocation eliminates
memory leaks by cyclically allocating objects out of a fixed-size buffer~\cite{NguyenR07}.

\noindent{\bf Data Structure Repair:} Data structure repair enables applications
to recover from data structure corruption errors~\cite{DemskyR03,DemskyR06}. It
enforces a data structure consistency specification. This specification
can be provided by a developer or automatically inferred from correct
program executions~\cite{DemskyEGMPR06}.

\noindent{\bf Self-Stabilizing Java:} Self-Stabilizing Java uses a type system
to ensure that the impact of any errors are eventually flushed from the
system, returning the system back to a consistent state~\cite{EomD12}.

\noindent{\bf Task Skipping and Loop Perforation:} 
Task skipping~\cite{Rinard06,Rinard07} discards task executions
and loop perforation~\cite{sidiroglou11} discards loop iterations. 
The result is a generated search space of programs with varying
performance and accuracy properties, some of which may be more
desirable than the original.  
In addition to enabling programs to survive tasks or loop iterations 
that trigger otherwise fatal errors, task skipping and loop perforation
can also improve performance.

\section{Conclusion}
\label{sec:conclusion}

A larger search space tends to contain correct patches for more defects, but
working with the larger space may increase the sparcity of the correct patches
and increase the chance of the correct patches being blocked by plausible but
incorrect patches. Understanding this tradeoff is important for the design of
future patch generation techniques. Our results show that, even with 
strong test suites, larger search spaces may decrease the ability of patch generation
systems to generate correct patches.

Systems such as ClearView and Prophet are successful, in part, because they leverage information
outside the validation test suite. To work
successfully with even richer search spaces, future systems will need to leverage
more sources of information that further enhance 
their ability to isolate the few correct patches
within larger sets of plausible but incorrect patches. 

%The difficulty of generating a search space rich enough to
%correct defects while still supporting an acceptably
%efficient search algorithm has significantly limited the ability of previous automatic patch
%generation systems to generate successful patches~\cite{LeGoues:2012,Weimer:2013,Kali}.
%{\name}'s novel combination of staged program repair, 
%parameterized transformation schemas, target value search,
%and condition synthesis highlight how
%a rich program repair search space coupled with an efficient
%search algorithm can enable successful automatic program repair. 

\section*{Acknowledgements}
We thank the anonymous reviewers for their insightful comments on early draft
of the paper. We thank Yu Lao for logistics support. This research was
supported by DARPA (Grant FA8650-11-C-7192 and FA8750-14-2-0242).

%This is the text of the appendix, if you need one.
\balance
\bibliographystyle{abbrv}
\bibliography{paper}

\newpage

%\twocolumn
\appendix

Tables~\ref{tab:app-first}-\ref{tab:app-last} present the detailed experimental
results. Each table presents the results of SPR or Prophet on one search space
configuration. The first column of the table presents the defect id. The second
column of the table presents the total number of candidate patch templates in
the search space. The third column presents the number of patch templates that
manipulate branch conditions. The fourth column presents the total number of
evaluated patch templates in 12 hours. The fifth column presents the total
number of evaluated condition patch templates during in 12 hours. The sixth
column presents the number of templates for which generate plausible patches.
The seventh column presents the number of condition templates for which
generate plausible patches. The eighth column presents the total number of
plausible patches the system finds in 12 hours. The ninth column presents the
number of plausible patches which manipulate branch conditions. The tenth
column presents the number of correct patches the system finds in 12 hours. The
eleventh column presents the rank of the template that generates the first
correct patch in the search space. The twelfth column presents the rank of the
template among plausible templates. The last column presents the rank of the
correct patch among all generated plausible patches.

\newpage

\begin{sidewaystable*}
\small
\centering
\begin{tabular}{|l|c|c|c|c|c|c|c|c|c|c|c|c|}
\hline
\multirow{3}{*}{\bf Defect} & \multicolumn{2}{|c|}{\bf Search Space} & \multicolumn{2}{|c|}{\bf Evaluated} & \multicolumn{2}{|c|}{\bf Plausible} & \multicolumn{2}{|c|}{\bf Plausible} & \multirow{3}{*}{\parbox[c]{1.2cm}{\bf Correct Patches}} & \multicolumn{2}{|c|}{\bf Correct} & {\bf Correct} \\
 &  \multicolumn{2}{|c|}{\bf Templates} & \multicolumn{2}{|c|}{\bf Templates} & \multicolumn{2}{|c|}{\bf Templates} & \multicolumn{2}{|c|}{\bf Patches} &     &  \multicolumn{2}{|c|}{\bf Template Rank} & {\bf Patch Rank} \\
\cline{2-3} \cline{4-5} \cline{6-7} \cline{8-9} \cline{11-12}
& {\bf All} & {\bf Cond.} &  {\bf All} & {\bf Cond.} & {\bf All} & {\bf Cond.} & {\bf All} & {\bf Cond.} &  & {\bf In Space} & {\bf In Plausible} & {\bf in Plausible} \\
\hline
{ \tiny gmp-14166-14167} & 10949 & 3394 & 10949 & 3394 & 33 & 28 & 111 & 106 & 0 &  - &  - &  - \\
\hline
{ \tiny libtiff-5b0217-3dfb33} & 83937 & 19298 & 52303 & 16953 & 208 & 147 & 211 & 150 & 0 &  - &  - &  - \\
\hline
{ \tiny libtiff-d13be7-ccadf4} & 184068 & 141307 & 63046 & 58498 & 1423 & 1423 & 1703 & 1423 & 1 & 1093 & 1 & 1 \\
\hline
{ \tiny lighttpd-2661-2662} & 34688 & 27541 & 34688 & 27541 & 70 & 8 & 73 & 11 & 0 &  - &  - &  - \\
\hline
{ \tiny lighttpd-1913-1914} & 32903 & 21707 & 32903 & 21707 & 4 & 4 & 4 & 4 & 0 &  - &  - &  - \\
\hline
\end{tabular}

\caption{Prophet-100 Statistics}
\label{tab:app-first}
\end{sidewaystable*}

\newpage

\begin{sidewaystable*}
\small
\centering
\begin{tabular}{|l|c|c|c|c|c|c|c|c|c|c|c|c|}
\hline
\multirow{3}{*}{\bf Defect} & \multicolumn{2}{|c|}{\bf Search Space} & \multicolumn{2}{|c|}{\bf Evaluated} & \multicolumn{2}{|c|}{\bf Plausible} & \multicolumn{2}{|c|}{\bf Plausible} & \multirow{3}{*}{\parbox[c]{1.2cm}{\bf Correct Patches}} & \multicolumn{2}{|c|}{\bf Correct} & {\bf Correct} \\
 &  \multicolumn{2}{|c|}{\bf Templates} & \multicolumn{2}{|c|}{\bf Templates} & \multicolumn{2}{|c|}{\bf Templates} & \multicolumn{2}{|c|}{\bf Patches} &     &  \multicolumn{2}{|c|}{\bf Template Rank} & {\bf Patch Rank} \\
\cline{2-3} \cline{4-5} \cline{6-7} \cline{8-9} \cline{11-12}
& {\bf All} & {\bf Cond.} &  {\bf All} & {\bf Cond.} & {\bf All} & {\bf Cond.} & {\bf All} & {\bf Cond.} &  & {\bf In Space} & {\bf In Plausible} & {\bf in Plausible} \\
\hline
{ \tiny gmp-14166-14167} & 10949 & 3394 & 10949 & 3394 & 29 & 24 & 56 & 51 & 0 &  - &  - &  - \\
\hline
{ \tiny libtiff-5b0217-3dfb33} & 83937 & 19298 & 1249 & 1112 & 146 & 146 & 3192 & 3192 & 0 &  - &  - &  - \\
\hline
{ \tiny libtiff-d13be7-ccadf4} & 184068 & 141307 & 3938 & 3896 & 116 & 116 & 4540 & 4362 & 2 & 1093 & 1 & 1 \\
\hline
{ \tiny lighttpd-2661-2662} & 34688 & 27541 & 34688 & 27541 & 78 & 16 & 109 & 47 & 0 &  - &  - &  - \\
\hline
{ \tiny lighttpd-1913-1914} & 32903 & 21707 & 32903 & 21707 & 4 & 4 & 12 & 12 & 0 &  - &  - &  - \\
\hline
\end{tabular}

\caption{Prophet-100-CExt Statistics}
\end{sidewaystable*}

\newpage

\begin{sidewaystable*}
\small
\centering
\begin{tabular}{|l|c|c|c|c|c|c|c|c|c|c|c|c|}
\hline
\multirow{3}{*}{\bf Defect} & \multicolumn{2}{|c|}{\bf Search Space} & \multicolumn{2}{|c|}{\bf Evaluated} & \multicolumn{2}{|c|}{\bf Plausible} & \multicolumn{2}{|c|}{\bf Plausible} & \multirow{3}{*}{\parbox[c]{1.2cm}{\bf Correct Patches}} & \multicolumn{2}{|c|}{\bf Correct} & {\bf Correct} \\
 &  \multicolumn{2}{|c|}{\bf Templates} & \multicolumn{2}{|c|}{\bf Templates} & \multicolumn{2}{|c|}{\bf Templates} & \multicolumn{2}{|c|}{\bf Patches} &     &  \multicolumn{2}{|c|}{\bf Template Rank} & {\bf Patch Rank} \\
\cline{2-3} \cline{4-5} \cline{6-7} \cline{8-9} \cline{11-12}
& {\bf All} & {\bf Cond.} &  {\bf All} & {\bf Cond.} & {\bf All} & {\bf Cond.} & {\bf All} & {\bf Cond.} &  & {\bf In Space} & {\bf In Plausible} & {\bf in Plausible} \\
\hline
{ \tiny gmp-14166-14167} & 12718 & 3394 & 12718 & 3394 & 33 & 28 & 111 & 106 & 0 &  - &  - &  - \\
\hline
{ \tiny libtiff-5b0217-3dfb33} & 88613 & 19298 & 53568 & 16729 & 210 & 147 & 213 & 150 & 0 &  - &  - &  - \\
\hline
{ \tiny libtiff-d13be7-ccadf4} & 186777 & 141307 & 60747 & 54316 & 1185 & 1185 & 1405 & 1185 & 1 & 1202 & 1 & 1 \\
\hline
{ \tiny lighttpd-2661-2662} & 35810 & 27541 & 35810 & 27541 & 66 & 4 & 69 & 7 & 0 &  - &  - &  - \\
\hline
{ \tiny lighttpd-1913-1914} & 34917 & 21707 & 34917 & 21707 & 4 & 4 & 4 & 4 & 0 &  - &  - &  - \\
\hline
\end{tabular}

\caption{Prophet-100-RExt Statistics}
\end{sidewaystable*}

\newpage

\begin{sidewaystable*}
\small
\centering
\begin{tabular}{|l|c|c|c|c|c|c|c|c|c|c|c|c|}
\hline
\multirow{3}{*}{\bf Defect} & \multicolumn{2}{|c|}{\bf Search Space} & \multicolumn{2}{|c|}{\bf Evaluated} & \multicolumn{2}{|c|}{\bf Plausible} & \multicolumn{2}{|c|}{\bf Plausible} & \multirow{3}{*}{\parbox[c]{1.2cm}{\bf Correct Patches}} & \multicolumn{2}{|c|}{\bf Correct} & {\bf Correct} \\
 &  \multicolumn{2}{|c|}{\bf Templates} & \multicolumn{2}{|c|}{\bf Templates} & \multicolumn{2}{|c|}{\bf Templates} & \multicolumn{2}{|c|}{\bf Patches} &     &  \multicolumn{2}{|c|}{\bf Template Rank} & {\bf Patch Rank} \\
\cline{2-3} \cline{4-5} \cline{6-7} \cline{8-9} \cline{11-12}
& {\bf All} & {\bf Cond.} &  {\bf All} & {\bf Cond.} & {\bf All} & {\bf Cond.} & {\bf All} & {\bf Cond.} &  & {\bf In Space} & {\bf In Plausible} & {\bf in Plausible} \\
\hline
{ \tiny gmp-14166-14167} & 12718 & 3394 & 12718 & 3394 & 29 & 24 & 56 & 51 & 0 &  - &  - &  - \\
\hline
{ \tiny libtiff-5b0217-3dfb33} & 88613 & 19298 & 1455 & 1325 & 146 & 146 & 3192 & 3192 & 0 &  - &  - &  - \\
\hline
{ \tiny libtiff-d13be7-ccadf4} & 186777 & 141307 & 4252 & 4207 & 115 & 115 & 4525 & 4347 & 2 & 1202 & 1 & 1 \\
\hline
{ \tiny lighttpd-2661-2662} & 35810 & 27541 & 35810 & 27541 & 74 & 12 & 97 & 35 & 0 &  - &  - &  - \\
\hline
{ \tiny lighttpd-1913-1914} & 34917 & 21707 & 34917 & 21707 & 4 & 4 & 12 & 12 & 0 &  - &  - &  - \\
\hline
\end{tabular}

\caption{Prophet-100-RExt-CExt Statistics}
\end{sidewaystable*}

\newpage

\begin{sidewaystable*}
\small
\centering
\begin{tabular}{|l|c|c|c|c|c|c|c|c|c|c|c|c|}
\hline
\multirow{3}{*}{\bf Defect} & \multicolumn{2}{|c|}{\bf Search Space} & \multicolumn{2}{|c|}{\bf Evaluated} & \multicolumn{2}{|c|}{\bf Plausible} & \multicolumn{2}{|c|}{\bf Plausible} & \multirow{3}{*}{\parbox[c]{1.2cm}{\bf Correct Patches}} & \multicolumn{2}{|c|}{\bf Correct} & {\bf Correct} \\
 &  \multicolumn{2}{|c|}{\bf Templates} & \multicolumn{2}{|c|}{\bf Templates} & \multicolumn{2}{|c|}{\bf Templates} & \multicolumn{2}{|c|}{\bf Patches} &     &  \multicolumn{2}{|c|}{\bf Template Rank} & {\bf Patch Rank} \\
\cline{2-3} \cline{4-5} \cline{6-7} \cline{8-9} \cline{11-12}
& {\bf All} & {\bf Cond.} &  {\bf All} & {\bf Cond.} & {\bf All} & {\bf Cond.} & {\bf All} & {\bf Cond.} &  & {\bf In Space} & {\bf In Plausible} & {\bf in Plausible} \\
\hline
{ \tiny gmp-14166-14167} & 39287 & 14161 & 39287 & 14161 & 33 & 28 & 111 & 106 & 0 &  - &  - &  - \\
\hline
{ \tiny libtiff-5b0217-3dfb33} & 221134 & 65017 & 64363 & 31834 & 247 & 149 & 250 & 152 & 0 & 50770 &  - &  - \\
\hline
{ \tiny libtiff-d13be7-ccadf4} & 296426 & 228659 & 70050 & 65731 & 1423 & 1423 & 1703 & 1423 & 1 & 1183 & 1 & 1 \\
\hline
{ \tiny lighttpd-2661-2662} & 120263 & 98028 & 79882 & 67844 & 48 & 2 & 50 & 4 & 0 &  - &  - &  - \\
\hline
{ \tiny lighttpd-1913-1914} & 68199 & 48133 & 37214 & 30701 & 58 & 58 & 58 & 58 & 0 &  - &  - &  - \\
\hline
\end{tabular}

\caption{Prophet-200 Statistics}
\end{sidewaystable*}
\newpage

\begin{sidewaystable*}
\small
\centering
\begin{tabular}{|l|c|c|c|c|c|c|c|c|c|c|c|c|}
\hline
\multirow{3}{*}{\bf Defect} & \multicolumn{2}{|c|}{\bf Search Space} & \multicolumn{2}{|c|}{\bf Evaluated} & \multicolumn{2}{|c|}{\bf Plausible} & \multicolumn{2}{|c|}{\bf Plausible} & \multirow{3}{*}{\parbox[c]{1.2cm}{\bf Correct Patches}} & \multicolumn{2}{|c|}{\bf Correct} & {\bf Correct} \\
 &  \multicolumn{2}{|c|}{\bf Templates} & \multicolumn{2}{|c|}{\bf Templates} & \multicolumn{2}{|c|}{\bf Templates} & \multicolumn{2}{|c|}{\bf Patches} &     &  \multicolumn{2}{|c|}{\bf Template Rank} & {\bf Patch Rank} \\
\cline{2-3} \cline{4-5} \cline{6-7} \cline{8-9} \cline{11-12}
& {\bf All} & {\bf Cond.} &  {\bf All} & {\bf Cond.} & {\bf All} & {\bf Cond.} & {\bf All} & {\bf Cond.} &  & {\bf In Space} & {\bf In Plausible} & {\bf in Plausible} \\
\hline
{ \tiny gmp-14166-14167} & 39287 & 14161 & 39287 & 14161 & 29 & 24 & 56 & 51 & 0 &  - &  - &  - \\
\hline
{ \tiny libtiff-5b0217-3dfb33} & 221134 & 65017 & 2054 & 1940 & 146 & 146 & 2776 & 2776 & 0 & 50770 &  - &  - \\
\hline
{ \tiny libtiff-d13be7-ccadf4} & 296426 & 228659 & 4679 & 4657 & 111 & 111 & 4308 & 4191 & 2 & 1183 & 1 & 1 \\
\hline
{ \tiny lighttpd-2661-2662} & 120263 & 98028 & 77400 & 65714 & 55 & 10 & 75 & 30 & 0 &  - &  - &  - \\
\hline
{ \tiny lighttpd-1913-1914} & 68199 & 48133 & 22474 & 19567 & 56 & 56 & 160 & 160 & 0 &  - &  - &  - \\
\hline
\end{tabular}

\caption{Prophet-200-CExt Statistics}
\end{sidewaystable*}
\newpage

\begin{sidewaystable*}
\small
\centering
\begin{tabular}{|l|c|c|c|c|c|c|c|c|c|c|c|c|}
\hline
\multirow{3}{*}{\bf Defect} & \multicolumn{2}{|c|}{\bf Search Space} & \multicolumn{2}{|c|}{\bf Evaluated} & \multicolumn{2}{|c|}{\bf Plausible} & \multicolumn{2}{|c|}{\bf Plausible} & \multirow{3}{*}{\parbox[c]{1.2cm}{\bf Correct Patches}} & \multicolumn{2}{|c|}{\bf Correct} & {\bf Correct} \\
 &  \multicolumn{2}{|c|}{\bf Templates} & \multicolumn{2}{|c|}{\bf Templates} & \multicolumn{2}{|c|}{\bf Templates} & \multicolumn{2}{|c|}{\bf Patches} &     &  \multicolumn{2}{|c|}{\bf Template Rank} & {\bf Patch Rank} \\
\cline{2-3} \cline{4-5} \cline{6-7} \cline{8-9} \cline{11-12}
& {\bf All} & {\bf Cond.} &  {\bf All} & {\bf Cond.} & {\bf All} & {\bf Cond.} & {\bf All} & {\bf Cond.} &  & {\bf In Space} & {\bf In Plausible} & {\bf in Plausible} \\
\hline
{ \tiny gmp-14166-14167} & 61171 & 14161 & 53629 & 13793 & 33 & 28 & 111 & 106 & 0 &  - &  - &  - \\
\hline
{ \tiny libtiff-5b0217-3dfb33} & 278857 & 65017 & 52657 & 22730 & 377 & 149 & 380 & 152 & 0 & 70740 &  - &  - \\
\hline
{ \tiny libtiff-d13be7-ccadf4} & 300745 & 228659 & 70497 & 63153 & 819 & 819 & 939 & 819 & 1 & 1321 & 1 & 1 \\
\hline
{ \tiny lighttpd-2661-2662} & 124449 & 98028 & 78470 & 64364 & 62 & 0 & 62 & 0 & 0 &  - &  - &  - \\
\hline
{ \tiny lighttpd-1913-1914} & 70712 & 48133 & 35590 & 27952 & 54 & 54 & 54 & 54 & 0 &  - &  - &  - \\
\hline
\end{tabular}

\caption{Prophet-200-RExt Statistics}
\end{sidewaystable*}
\newpage

\begin{sidewaystable*}
\small
\centering
\begin{tabular}{|l|c|c|c|c|c|c|c|c|c|c|c|c|}
\hline
\multirow{3}{*}{\bf Defect} & \multicolumn{2}{|c|}{\bf Search Space} & \multicolumn{2}{|c|}{\bf Evaluated} & \multicolumn{2}{|c|}{\bf Plausible} & \multicolumn{2}{|c|}{\bf Plausible} & \multirow{3}{*}{\parbox[c]{1.2cm}{\bf Correct Patches}} & \multicolumn{2}{|c|}{\bf Correct} & {\bf Correct} \\
 &  \multicolumn{2}{|c|}{\bf Templates} & \multicolumn{2}{|c|}{\bf Templates} & \multicolumn{2}{|c|}{\bf Templates} & \multicolumn{2}{|c|}{\bf Patches} &     &  \multicolumn{2}{|c|}{\bf Template Rank} & {\bf Patch Rank} \\
\cline{2-3} \cline{4-5} \cline{6-7} \cline{8-9} \cline{11-12}
& {\bf All} & {\bf Cond.} &  {\bf All} & {\bf Cond.} & {\bf All} & {\bf Cond.} & {\bf All} & {\bf Cond.} &  & {\bf In Space} & {\bf In Plausible} & {\bf in Plausible} \\
\hline
{ \tiny gmp-14166-14167} & 61171 & 14161 & 55865 & 14057 & 29 & 24 & 56 & 51 & 0 &  - &  - &  - \\
\hline
{ \tiny libtiff-5b0217-3dfb33} & 278857 & 65017 & 2032 & 1923 & 146 & 146 & 2806 & 2806 & 0 & 70740 &  - &  - \\
\hline
{ \tiny libtiff-d13be7-ccadf4} & 300745 & 228659 & 4825 & 4803 & 109 & 109 & 4279 & 4162 & 2 & 1321 & 1 & 1 \\
\hline
{ \tiny lighttpd-2661-2662} & 124449 & 98028 & 77686 & 64125 & 64 & 10 & 81 & 27 & 0 &  - &  - &  - \\
\hline
{ \tiny lighttpd-1913-1914} & 70712 & 48133 & 23341 & 19172 & 55 & 55 & 155 & 155 & 0 &  - &  - &  - \\
\hline
\end{tabular}

\caption{Prophet-200-RExt-CExt Statistics}
\end{sidewaystable*}
\newpage

\begin{sidewaystable*}
\small
\centering
\begin{tabular}{|l|c|c|c|c|c|c|c|c|c|c|c|c|}
\hline
\multirow{3}{*}{\bf Defect} & \multicolumn{2}{|c|}{\bf Search Space} & \multicolumn{2}{|c|}{\bf Evaluated} & \multicolumn{2}{|c|}{\bf Plausible} & \multicolumn{2}{|c|}{\bf Plausible} & \multirow{3}{*}{\parbox[c]{1.2cm}{\bf Correct Patches}} & \multicolumn{2}{|c|}{\bf Correct} & {\bf Correct} \\
 &  \multicolumn{2}{|c|}{\bf Templates} & \multicolumn{2}{|c|}{\bf Templates} & \multicolumn{2}{|c|}{\bf Templates} & \multicolumn{2}{|c|}{\bf Patches} &     &  \multicolumn{2}{|c|}{\bf Template Rank} & {\bf Patch Rank} \\
\cline{2-3} \cline{4-5} \cline{6-7} \cline{8-9} \cline{11-12}
& {\bf All} & {\bf Cond.} &  {\bf All} & {\bf Cond.} & {\bf All} & {\bf Cond.} & {\bf All} & {\bf Cond.} &  & {\bf In Space} & {\bf In Plausible} & {\bf in Plausible} \\
\hline
{ \tiny gmp-14166-14167} & 59327 & 22055 & 49749 & 21743 & 33 & 28 & 111 & 106 & 0 &  - &  - &  - \\
\hline
{ \tiny libtiff-5b0217-3dfb33} & 276142 & 85251 & 64841 & 32860 & 225 & 149 & 228 & 152 & 0 & 52950 &  - &  - \\
\hline
{ \tiny libtiff-d13be7-ccadf4} & 568172 & 432083 & 76111 & 71854 & 1423 & 1423 & 1703 & 1423 & 1 & 1453 & 1 & 1 \\
\hline
{ \tiny lighttpd-2661-2662} & 191579 & 159862 & 92356 & 80355 & 48 & 3 & 51 & 6 & 0 &  - &  - &  - \\
\hline
{ \tiny lighttpd-1913-1914} & 159739 & 114667 & 36479 & 30861 & 63 & 63 & 63 & 63 & 0 &  - &  - &  - \\
\hline
\end{tabular}

\caption{Prophet-300 Statistics}
\end{sidewaystable*}

\begin{sidewaystable*}
\small
\centering
\begin{tabular}{|l|c|c|c|c|c|c|c|c|c|c|c|c|}
\hline
\multirow{3}{*}{\bf Defect} & \multicolumn{2}{|c|}{\bf Search Space} & \multicolumn{2}{|c|}{\bf Evaluated} & \multicolumn{2}{|c|}{\bf Plausible} & \multicolumn{2}{|c|}{\bf Plausible} & \multirow{3}{*}{\parbox[c]{1.2cm}{\bf Correct Patches}} & \multicolumn{2}{|c|}{\bf Correct} & {\bf Correct} \\
 &  \multicolumn{2}{|c|}{\bf Templates} & \multicolumn{2}{|c|}{\bf Templates} & \multicolumn{2}{|c|}{\bf Templates} & \multicolumn{2}{|c|}{\bf Patches} &     &  \multicolumn{2}{|c|}{\bf Template Rank} & {\bf Patch Rank} \\
\cline{2-3} \cline{4-5} \cline{6-7} \cline{8-9} \cline{11-12}
& {\bf All} & {\bf Cond.} &  {\bf All} & {\bf Cond.} & {\bf All} & {\bf Cond.} & {\bf All} & {\bf Cond.} &  & {\bf In Space} & {\bf In Plausible} & {\bf in Plausible} \\
\hline
{ \tiny gmp-14166-14167} & 59327 & 22055 & 51701 & 21934 & 29 & 24 & 56 & 51 & 0 &  - &  - &  - \\
\hline
{ \tiny libtiff-5b0217-3dfb33} & 276142 & 85251 & 2433 & 2326 & 146 & 146 & 2794 & 2794 & 0 & 52950 &  - &  - \\
\hline
{ \tiny libtiff-d13be7-ccadf4} & 568172 & 432083 & 10788 & 10704 & 108 & 108 & 4237 & 4120 & 2 & 1453 & 1 & 1 \\
\hline
{ \tiny lighttpd-2661-2662} & 191579 & 159862 & 90638 & 78887 & 59 & 14 & 80 & 35 & 0 &  - &  - &  - \\
\hline
{ \tiny lighttpd-1913-1914} & 159739 & 114667 & 22555 & 20308 & 57 & 57 & 160 & 160 & 0 & 42256 &  - &  - \\
\hline
\end{tabular}

\caption{Prophet-300-CExt Statistics}
\end{sidewaystable*}

\begin{sidewaystable*}
\small
\centering
\begin{tabular}{|l|c|c|c|c|c|c|c|c|c|c|c|c|}
\hline
\multirow{3}{*}{\bf Defect} & \multicolumn{2}{|c|}{\bf Search Space} & \multicolumn{2}{|c|}{\bf Evaluated} & \multicolumn{2}{|c|}{\bf Plausible} & \multicolumn{2}{|c|}{\bf Plausible} & \multirow{3}{*}{\parbox[c]{1.2cm}{\bf Correct Patches}} & \multicolumn{2}{|c|}{\bf Correct} & {\bf Correct} \\
 &  \multicolumn{2}{|c|}{\bf Templates} & \multicolumn{2}{|c|}{\bf Templates} & \multicolumn{2}{|c|}{\bf Templates} & \multicolumn{2}{|c|}{\bf Patches} &     &  \multicolumn{2}{|c|}{\bf Template Rank} & {\bf Patch Rank} \\
\cline{2-3} \cline{4-5} \cline{6-7} \cline{8-9} \cline{11-12}
& {\bf All} & {\bf Cond.} &  {\bf All} & {\bf Cond.} & {\bf All} & {\bf Cond.} & {\bf All} & {\bf Cond.} &  & {\bf In Space} & {\bf In Plausible} & {\bf in Plausible} \\
\hline
{ \tiny gmp-14166-14167} & 95751 & 22055 & 43517 & 13623 & 31 & 28 & 109 & 106 & 0 & 62583 &  - &  - \\
\hline
{ \tiny libtiff-5b0217-3dfb33} & 336808 & 85251 & 53883 & 23725 & 303 & 149 & 306 & 152 & 0 & 72938 &  - &  - \\
\hline
{ \tiny libtiff-d13be7-ccadf4} & 573714 & 432083 & 76021 & 68438 & 556 & 556 & 609 & 556 & 1 & 1608 & 1 & 1 \\
\hline
{ \tiny lighttpd-2661-2662} & 197676 & 159862 & 90518 & 76321 & 58 & 4 & 61 & 7 & 0 &  - &  - &  - \\
\hline
{ \tiny lighttpd-1913-1914} & 168946 & 114667 & 36761 & 29588 & 60 & 60 & 60 & 60 & 0 &  - &  - &  - \\
\hline
\end{tabular}

\caption{Prophet-300-RExt Statistics}
\end{sidewaystable*}

\begin{sidewaystable*}
\small
\centering
\begin{tabular}{|l|c|c|c|c|c|c|c|c|c|c|c|c|}
\hline
\multirow{3}{*}{\bf Defect} & \multicolumn{2}{|c|}{\bf Search Space} & \multicolumn{2}{|c|}{\bf Evaluated} & \multicolumn{2}{|c|}{\bf Plausible} & \multicolumn{2}{|c|}{\bf Plausible} & \multirow{3}{*}{\parbox[c]{1.2cm}{\bf Correct Patches}} & \multicolumn{2}{|c|}{\bf Correct} & {\bf Correct} \\
 &  \multicolumn{2}{|c|}{\bf Templates} & \multicolumn{2}{|c|}{\bf Templates} & \multicolumn{2}{|c|}{\bf Templates} & \multicolumn{2}{|c|}{\bf Patches} &     &  \multicolumn{2}{|c|}{\bf Template Rank} & {\bf Patch Rank} \\
\cline{2-3} \cline{4-5} \cline{6-7} \cline{8-9} \cline{11-12}
& {\bf All} & {\bf Cond.} &  {\bf All} & {\bf Cond.} & {\bf All} & {\bf Cond.} & {\bf All} & {\bf Cond.} &  & {\bf In Space} & {\bf In Plausible} & {\bf in Plausible} \\
\hline
{ \tiny gmp-14166-14167} & 95751 & 22055 & 45224 & 14092 & 27 & 24 & 54 & 51 & 0 & 62583 &  - &  - \\
\hline
{ \tiny libtiff-5b0217-3dfb33} & 336808 & 85251 & 2369 & 2261 & 146 & 146 & 2848 & 2848 & 0 & 72938 &  - &  - \\
\hline
{ \tiny libtiff-d13be7-ccadf4} & 573714 & 432083 & 11655 & 11585 & 107 & 107 & 4205 & 4088 & 2 & 1608 & 1 & 1 \\
\hline
{ \tiny lighttpd-2661-2662} & 197676 & 159862 & 90748 & 76551 & 57 & 3 & 59 & 5 & 0 &  - &  - &  - \\
\hline
{ \tiny lighttpd-1913-1914} & 168946 & 114667 & 23858 & 20291 & 54 & 54 & 149 & 149 & 0 & 46828 &  - &  - \\
\hline
\end{tabular}

\caption{Prophet-300-RExt-CExt Statistics}
\end{sidewaystable*}

\begin{sidewaystable*}
\small
\centering
\begin{tabular}{|l|c|c|c|c|c|c|c|c|c|c|c|c|}
\hline
\multirow{3}{*}{\bf Defect} & \multicolumn{2}{|c|}{\bf Search Space} & \multicolumn{2}{|c|}{\bf Evaluated} & \multicolumn{2}{|c|}{\bf Plausible} & \multicolumn{2}{|c|}{\bf Plausible} & \multirow{3}{*}{\parbox[c]{1.2cm}{\bf Correct Patches}} & \multicolumn{2}{|c|}{\bf Correct} & {\bf Correct} \\
 &  \multicolumn{2}{|c|}{\bf Templates} & \multicolumn{2}{|c|}{\bf Templates} & \multicolumn{2}{|c|}{\bf Templates} & \multicolumn{2}{|c|}{\bf Patches} &     &  \multicolumn{2}{|c|}{\bf Template Rank} & {\bf Patch Rank} \\
\cline{2-3} \cline{4-5} \cline{6-7} \cline{8-9} \cline{11-12}
& {\bf All} & {\bf Cond.} &  {\bf All} & {\bf Cond.} & {\bf All} & {\bf Cond.} & {\bf All} & {\bf Cond.} &  & {\bf In Space} & {\bf In Plausible} & {\bf in Plausible} \\
\hline
{ \tiny gmp-14166-14167} & 448318 & 141182 & 43569 & 22779 & 31 & 28 & 109 & 106 & 0 &  - &  - &  - \\
\hline
{ \tiny libtiff-5b0217-3dfb33} & 1941177 & 1352092 & 83094 & 59078 & 182 & 149 & 185 & 152 & 0 & 91329 &  - &  - \\
\hline
{ \tiny libtiff-d13be7-ccadf4} & 2353240 & 1771135 & 113514 & 106810 & 416 & 416 & 431 & 416 & 1 & 2820 & 1 & 1 \\
\hline
{ \tiny lighttpd-2661-2662} & 1269307 & 1016413 & 120469 & 107039 & 46 & 10 & 54 & 18 & 0 & 1197010 &  - &  - \\
\hline
{ \tiny lighttpd-1913-1914} & 1183286 & 935121 & 59987 & 56385 & 51 & 51 & 51 & 51 & 0 &  - &  - &  - \\
\hline
\end{tabular}

\caption{Prophet-2000 Statistics}
\end{sidewaystable*}

\begin{sidewaystable*}
\small
\centering
\begin{tabular}{|l|c|c|c|c|c|c|c|c|c|c|c|c|}
\hline
\multirow{3}{*}{\bf Defect} & \multicolumn{2}{|c|}{\bf Search Space} & \multicolumn{2}{|c|}{\bf Evaluated} & \multicolumn{2}{|c|}{\bf Plausible} & \multicolumn{2}{|c|}{\bf Plausible} & \multirow{3}{*}{\parbox[c]{1.2cm}{\bf Correct Patches}} & \multicolumn{2}{|c|}{\bf Correct} & {\bf Correct} \\
 &  \multicolumn{2}{|c|}{\bf Templates} & \multicolumn{2}{|c|}{\bf Templates} & \multicolumn{2}{|c|}{\bf Templates} & \multicolumn{2}{|c|}{\bf Patches} &     &  \multicolumn{2}{|c|}{\bf Template Rank} & {\bf Patch Rank} \\
\cline{2-3} \cline{4-5} \cline{6-7} \cline{8-9} \cline{11-12}
& {\bf All} & {\bf Cond.} &  {\bf All} & {\bf Cond.} & {\bf All} & {\bf Cond.} & {\bf All} & {\bf Cond.} &  & {\bf In Space} & {\bf In Plausible} & {\bf in Plausible} \\
\hline
{ \tiny gmp-14166-14167} & 448318 & 141182 & 45987 & 23728 & 27 & 24 & 54 & 51 & 0 &  - &  - &  - \\
\hline
{ \tiny libtiff-5b0217-3dfb33} & 1941177 & 1352092 & 21932 & 21903 & 111 & 111 & 2276 & 2276 & 0 & 91329 &  - &  - \\
\hline
{ \tiny libtiff-d13be7-ccadf4} & 2353240 & 1771135 & 34658 & 34608 & 103 & 103 & 3559 & 3446 & 2 & 2820 & 1 & 1 \\
\hline
{ \tiny lighttpd-2661-2662} & 1269307 & 1016413 & 115409 & 103056 & 57 & 23 & 92 & 58 & 0 & 1197010 &  - &  - \\
\hline
{ \tiny lighttpd-1913-1914} & 1183286 & 935121 & 57046 & 54412 & 40 & 40 & 108 & 108 & 0 & 81194 &  - &  - \\
\hline
\end{tabular}

\caption{Prophet-2000-CExt Statistics}
\end{sidewaystable*}

\begin{sidewaystable*}
\small
\centering
\begin{tabular}{|l|c|c|c|c|c|c|c|c|c|c|c|c|}
\hline
\multirow{3}{*}{\bf Defect} & \multicolumn{2}{|c|}{\bf Search Space} & \multicolumn{2}{|c|}{\bf Evaluated} & \multicolumn{2}{|c|}{\bf Plausible} & \multicolumn{2}{|c|}{\bf Plausible} & \multirow{3}{*}{\parbox[c]{1.2cm}{\bf Correct Patches}} & \multicolumn{2}{|c|}{\bf Correct} & {\bf Correct} \\
 &  \multicolumn{2}{|c|}{\bf Templates} & \multicolumn{2}{|c|}{\bf Templates} & \multicolumn{2}{|c|}{\bf Templates} & \multicolumn{2}{|c|}{\bf Patches} &     &  \multicolumn{2}{|c|}{\bf Template Rank} & {\bf Patch Rank} \\
\cline{2-3} \cline{4-5} \cline{6-7} \cline{8-9} \cline{11-12}
& {\bf All} & {\bf Cond.} &  {\bf All} & {\bf Cond.} & {\bf All} & {\bf Cond.} & {\bf All} & {\bf Cond.} &  & {\bf In Space} & {\bf In Plausible} & {\bf in Plausible} \\
\hline
{ \tiny gmp-14166-14167} & 780425 & 141182 & 35880 & 14864 & 30 & 28 & 108 & 106 & 0 & 87034 &  - &  - \\
\hline
{ \tiny libtiff-5b0217-3dfb33} & 2027218 & 1352092 & 71298 & 51369 & 154 & 149 & 157 & 152 & 0 & 112099 &  - &  - \\
\hline
{ \tiny libtiff-d13be7-ccadf4} & 2414317 & 1771135 & 113033 & 104624 & 416 & 416 & 431 & 416 & 1 & 4324 & 1 & 1 \\
\hline
{ \tiny lighttpd-2661-2662} & 1290186 & 1016413 & 110266 & 95506 & 60 & 11 & 68 & 19 & 0 & 1217473 &  - &  - \\
\hline
{ \tiny lighttpd-1913-1914} & 1206699 & 935121 & 60268 & 55919 & 48 & 48 & 48 & 48 & 0 &  - &  - &  - \\
\hline
\end{tabular}

\caption{Prophet-2000-RExt Statistics}
\end{sidewaystable*}

\begin{sidewaystable*}
\small
\centering
\begin{tabular}{|l|c|c|c|c|c|c|c|c|c|c|c|c|}
\hline
\multirow{3}{*}{\bf Defect} & \multicolumn{2}{|c|}{\bf Search Space} & \multicolumn{2}{|c|}{\bf Evaluated} & \multicolumn{2}{|c|}{\bf Plausible} & \multicolumn{2}{|c|}{\bf Plausible} & \multirow{3}{*}{\parbox[c]{1.2cm}{\bf Correct Patches}} & \multicolumn{2}{|c|}{\bf Correct} & {\bf Correct} \\
 &  \multicolumn{2}{|c|}{\bf Templates} & \multicolumn{2}{|c|}{\bf Templates} & \multicolumn{2}{|c|}{\bf Templates} & \multicolumn{2}{|c|}{\bf Patches} &     &  \multicolumn{2}{|c|}{\bf Template Rank} & {\bf Patch Rank} \\
\cline{2-3} \cline{4-5} \cline{6-7} \cline{8-9} \cline{11-12}
& {\bf All} & {\bf Cond.} &  {\bf All} & {\bf Cond.} & {\bf All} & {\bf Cond.} & {\bf All} & {\bf Cond.} &  & {\bf In Space} & {\bf In Plausible} & {\bf in Plausible} \\
\hline
{ \tiny gmp-14166-14167} & 780425 & 141182 & 39259 & 15947 & 27 & 24 & 54 & 51 & 0 & 87034 &  - &  - \\
\hline
{ \tiny libtiff-5b0217-3dfb33} & 2027218 & 1352092 & 22673 & 22656 & 111 & 111 & 2178 & 2178 & 0 & 112099 &  - &  - \\
\hline
{ \tiny libtiff-d13be7-ccadf4} & 2414317 & 1771135 & 37308 & 37266 & 93 & 93 & 3371 & 3258 & 2 & 4324 & 1 & 1 \\
\hline
{ \tiny lighttpd-2661-2662} & 1290186 & 1016413 & 104869 & 90483 & 69 & 23 & 102 & 56 & 0 & 1217473 &  - &  - \\
\hline
{ \tiny lighttpd-1913-1914} & 1206699 & 935121 & 57257 & 54013 & 42 & 42 & 109 & 109 & 0 & 86441 &  - &  - \\
\hline
\end{tabular}

\caption{Prophet-2000-RExt-CExt Statistics}
\end{sidewaystable*}

\clearpage
\begin{sidewaystable*}
\small
\centering
\begin{tabular}{|l|c|c|c|c|c|c|c|c|c|c|c|c|}
\hline
\multirow{3}{*}{\bf Defect} & \multicolumn{2}{|c|}{\bf Search Space} & \multicolumn{2}{|c|}{\bf Evaluated} & \multicolumn{2}{|c|}{\bf Plausible} & \multicolumn{2}{|c|}{\bf Plausible} & \multirow{3}{*}{\parbox[c]{1.2cm}{\bf Correct Patches}} & \multicolumn{2}{|c|}{\bf Correct} & {\bf Correct} \\
 &  \multicolumn{2}{|c|}{\bf Templates} & \multicolumn{2}{|c|}{\bf Templates} & \multicolumn{2}{|c|}{\bf Templates} & \multicolumn{2}{|c|}{\bf Patches} &     &  \multicolumn{2}{|c|}{\bf Template Rank} & {\bf Patch Rank} \\
\cline{2-3} \cline{4-5} \cline{6-7} \cline{8-9} \cline{11-12}
& {\bf All} & {\bf Cond.} &  {\bf All} & {\bf Cond.} & {\bf All} & {\bf Cond.} & {\bf All} & {\bf Cond.} &  & {\bf In Space} & {\bf In Plausible} & {\bf in Plausible} \\
\hline
{ \tiny gmp-14166-14167} & 10949 & 3394 & 10949 & 3394 & 33 & 28 & 111 & 106 & 0 &  - &  - &  - \\
\hline
{ \tiny libtiff-5b0217-3dfb33} & 83937 & 19298 & 64840 & 19298 & 174 & 147 & 177 & 150 & 0 &  - &  - &  - \\
\hline
{ \tiny libtiff-d13be7-ccadf4} & 184068 & 141307 & 45961 & 45374 & 1723 & 1723 & 2003 & 1723 & 1 & 2408 & 3 & 3 \\
\hline
{ \tiny lighttpd-2661-2662} & 34688 & 27541 & 34688 & 27541 & 70 & 8 & 73 & 11 & 0 &  - &  - &  - \\
\hline
{ \tiny lighttpd-1913-1914} & 32903 & 21707 & 32903 & 21707 & 4 & 4 & 4 & 4 & 0 &  - &  - &  - \\
\hline
\end{tabular}

\caption{SPR-100 Statistics}
\end{sidewaystable*}

\begin{sidewaystable*}
\small
\centering
\begin{tabular}{|l|c|c|c|c|c|c|c|c|c|c|c|c|}
\hline
\multirow{3}{*}{\bf Defect} & \multicolumn{2}{|c|}{\bf Search Space} & \multicolumn{2}{|c|}{\bf Evaluated} & \multicolumn{2}{|c|}{\bf Plausible} & \multicolumn{2}{|c|}{\bf Plausible} & \multirow{3}{*}{\parbox[c]{1.2cm}{\bf Correct Patches}} & \multicolumn{2}{|c|}{\bf Correct} & {\bf Correct} \\
 &  \multicolumn{2}{|c|}{\bf Templates} & \multicolumn{2}{|c|}{\bf Templates} & \multicolumn{2}{|c|}{\bf Templates} & \multicolumn{2}{|c|}{\bf Patches} &     &  \multicolumn{2}{|c|}{\bf Template Rank} & {\bf Patch Rank} \\
\cline{2-3} \cline{4-5} \cline{6-7} \cline{8-9} \cline{11-12}
& {\bf All} & {\bf Cond.} &  {\bf All} & {\bf Cond.} & {\bf All} & {\bf Cond.} & {\bf All} & {\bf Cond.} &  & {\bf In Space} & {\bf In Plausible} & {\bf in Plausible} \\
\hline
{ \tiny gmp-14166-14167} & 10949 & 3394 & 10949 & 3394 & 29 & 24 & 56 & 51 & 0 &  - &  - &  - \\
\hline
{ \tiny libtiff-5b0217-3dfb33} & 83937 & 19298 & 13487 & 5683 & 162 & 162 & 2879 & 2879 & 0 &  - &  - &  - \\
\hline
{ \tiny libtiff-d13be7-ccadf4} & 184068 & 141307 & 5269 & 5269 & 247 & 247 & 4450 & 4450 & 2 & 2408 & 5 & 104 \\
\hline
{ \tiny lighttpd-2661-2662} & 34688 & 27541 & 34688 & 27541 & 78 & 16 & 109 & 47 & 0 &  - &  - &  - \\
\hline
{ \tiny lighttpd-1913-1914} & 32903 & 21707 & 32903 & 21707 & 4 & 4 & 12 & 12 & 0 &  - &  - &  - \\
\hline
\end{tabular}

\caption{SPR-100-CExt Statistics}
\end{sidewaystable*}

\begin{sidewaystable*}
\small
\centering
\begin{tabular}{|l|c|c|c|c|c|c|c|c|c|c|c|c|}
\hline
\multirow{3}{*}{\bf Defect} & \multicolumn{2}{|c|}{\bf Search Space} & \multicolumn{2}{|c|}{\bf Evaluated} & \multicolumn{2}{|c|}{\bf Plausible} & \multicolumn{2}{|c|}{\bf Plausible} & \multirow{3}{*}{\parbox[c]{1.2cm}{\bf Correct Patches}} & \multicolumn{2}{|c|}{\bf Correct} & {\bf Correct} \\
 &  \multicolumn{2}{|c|}{\bf Templates} & \multicolumn{2}{|c|}{\bf Templates} & \multicolumn{2}{|c|}{\bf Templates} & \multicolumn{2}{|c|}{\bf Patches} &     &  \multicolumn{2}{|c|}{\bf Template Rank} & {\bf Patch Rank} \\
\cline{2-3} \cline{4-5} \cline{6-7} \cline{8-9} \cline{11-12}
& {\bf All} & {\bf Cond.} &  {\bf All} & {\bf Cond.} & {\bf All} & {\bf Cond.} & {\bf All} & {\bf Cond.} &  & {\bf In Space} & {\bf In Plausible} & {\bf in Plausible} \\
\hline
{ \tiny gmp-14166-14167} & 12718 & 3394 & 12718 & 3394 & 33 & 28 & 111 & 106 & 0 &  - &  - &  - \\
\hline
{ \tiny libtiff-5b0217-3dfb33} & 88613 & 19298 & 68340 & 19298 & 174 & 147 & 177 & 150 & 0 &  - &  - &  - \\
\hline
{ \tiny libtiff-d13be7-ccadf4} & 186777 & 141307 & 160494 & 141307 & 1723 & 1723 & 2003 & 1723 & 1 & 2117 & 3 & 3 \\
\hline
{ \tiny lighttpd-2661-2662} & 35810 & 27541 & 35810 & 27541 & 66 & 4 & 66 & 4 & 0 &  - &  - &  - \\
\hline
{ \tiny lighttpd-1913-1914} & 34917 & 21707 & 34917 & 21707 & 4 & 4 & 4 & 4 & 0 &  - &  - &  - \\
\hline
\end{tabular}

\caption{SPR-100-RExt Statistics}
\end{sidewaystable*}

\begin{sidewaystable*}
\small
\centering
\begin{tabular}{|l|c|c|c|c|c|c|c|c|c|c|c|c|}
\hline
\multirow{3}{*}{\bf Defect} & \multicolumn{2}{|c|}{\bf Search Space} & \multicolumn{2}{|c|}{\bf Evaluated} & \multicolumn{2}{|c|}{\bf Plausible} & \multicolumn{2}{|c|}{\bf Plausible} & \multirow{3}{*}{\parbox[c]{1.2cm}{\bf Correct Patches}} & \multicolumn{2}{|c|}{\bf Correct} & {\bf Correct} \\
 &  \multicolumn{2}{|c|}{\bf Templates} & \multicolumn{2}{|c|}{\bf Templates} & \multicolumn{2}{|c|}{\bf Templates} & \multicolumn{2}{|c|}{\bf Patches} &     &  \multicolumn{2}{|c|}{\bf Template Rank} & {\bf Patch Rank} \\
\cline{2-3} \cline{4-5} \cline{6-7} \cline{8-9} \cline{11-12}
& {\bf All} & {\bf Cond.} &  {\bf All} & {\bf Cond.} & {\bf All} & {\bf Cond.} & {\bf All} & {\bf Cond.} &  & {\bf In Space} & {\bf In Plausible} & {\bf in Plausible} \\
\hline
{ \tiny gmp-14166-14167} & 12718 & 3394 & 12718 & 3394 & 29 & 24 & 56 & 51 & 0 &  - &  - &  - \\
\hline
{ \tiny libtiff-5b0217-3dfb33} & 88613 & 19298 & 10264 & 5607 & 136 & 136 & 2721 & 2721 & 0 &  - &  - &  - \\
\hline
{ \tiny libtiff-d13be7-ccadf4} & 186777 & 141307 & 5269 & 5269 & 232 & 232 & 4350 & 4350 & 2 & 2117 & 5 & 104 \\
\hline
{ \tiny lighttpd-2661-2662} & 35810 & 27541 & 35810 & 27541 & 74 & 12 & 97 & 35 & 0 &  - &  - &  - \\
\hline
{ \tiny lighttpd-1913-1914} & 34917 & 21707 & 34917 & 21707 & 4 & 4 & 12 & 12 & 0 &  - &  - &  - \\
\hline
\end{tabular}

\caption{SPR-100-RExt-CExt Statistics}
\end{sidewaystable*}

\begin{sidewaystable*}
\small
\centering
\begin{tabular}{|l|c|c|c|c|c|c|c|c|c|c|c|c|}
\hline
\multirow{3}{*}{\bf Defect} & \multicolumn{2}{|c|}{\bf Search Space} & \multicolumn{2}{|c|}{\bf Evaluated} & \multicolumn{2}{|c|}{\bf Plausible} & \multicolumn{2}{|c|}{\bf Plausible} & \multirow{3}{*}{\parbox[c]{1.2cm}{\bf Correct Patches}} & \multicolumn{2}{|c|}{\bf Correct} & {\bf Correct} \\
 &  \multicolumn{2}{|c|}{\bf Templates} & \multicolumn{2}{|c|}{\bf Templates} & \multicolumn{2}{|c|}{\bf Templates} & \multicolumn{2}{|c|}{\bf Patches} &     &  \multicolumn{2}{|c|}{\bf Template Rank} & {\bf Patch Rank} \\
\cline{2-3} \cline{4-5} \cline{6-7} \cline{8-9} \cline{11-12}
& {\bf All} & {\bf Cond.} &  {\bf All} & {\bf Cond.} & {\bf All} & {\bf Cond.} & {\bf All} & {\bf Cond.} &  & {\bf In Space} & {\bf In Plausible} & {\bf in Plausible} \\
\hline
{ \tiny gmp-14166-14167} & 39287 & 14161 & 33154 & 14161 & 34 & 28 & 112 & 106 & 0 &  - &  - &  - \\
\hline
{ \tiny libtiff-5b0217-3dfb33} & 221134 & 65017 & 107534 & 65017 & 237 & 149 & 240 & 152 & 1 & 56644 & 208 & 211 \\
\hline
{ \tiny libtiff-d13be7-ccadf4} & 296426 & 228659 & 230917 & 214697 & 1723 & 1723 & 2003 & 1723 & 1 & 372 & 3 & 3 \\
\hline
{ \tiny lighttpd-2661-2662} & 120263 & 98028 & 105423 & 98028 & 52 & 4 & 55 & 7 & 0 &  - &  - &  - \\
\hline
{ \tiny lighttpd-1913-1914} & 68199 & 48133 & 45191 & 40674 & 55 & 55 & 55 & 55 & 0 &  - &  - &  - \\
\hline
\end{tabular}

\caption{SPR-200 Statistics}
\end{sidewaystable*}

\begin{sidewaystable*}
\small
\centering
\begin{tabular}{|l|c|c|c|c|c|c|c|c|c|c|c|c|}
\hline
\multirow{3}{*}{\bf Defect} & \multicolumn{2}{|c|}{\bf Search Space} & \multicolumn{2}{|c|}{\bf Evaluated} & \multicolumn{2}{|c|}{\bf Plausible} & \multicolumn{2}{|c|}{\bf Plausible} & \multirow{3}{*}{\parbox[c]{1.2cm}{\bf Correct Patches}} & \multicolumn{2}{|c|}{\bf Correct} & {\bf Correct} \\
 &  \multicolumn{2}{|c|}{\bf Templates} & \multicolumn{2}{|c|}{\bf Templates} & \multicolumn{2}{|c|}{\bf Templates} & \multicolumn{2}{|c|}{\bf Patches} &     &  \multicolumn{2}{|c|}{\bf Template Rank} & {\bf Patch Rank} \\
\cline{2-3} \cline{4-5} \cline{6-7} \cline{8-9} \cline{11-12}
& {\bf All} & {\bf Cond.} &  {\bf All} & {\bf Cond.} & {\bf All} & {\bf Cond.} & {\bf All} & {\bf Cond.} &  & {\bf In Space} & {\bf In Plausible} & {\bf in Plausible} \\
\hline
{ \tiny gmp-14166-14167} & 39287 & 14161 & 34474 & 14161 & 30 & 24 & 57 & 51 & 0 &  - &  - &  - \\
\hline
{ \tiny libtiff-5b0217-3dfb33} & 221134 & 65017 & 13644 & 8420 & 111 & 111 & 2604 & 2604 & 0 & 56644 &  - &  - \\
\hline
{ \tiny libtiff-d13be7-ccadf4} & 296426 & 228659 & 13486 & 13486 & 225 & 225 & 4240 & 4240 & 2 & 372 & 5 & 104 \\
\hline
{ \tiny lighttpd-2661-2662} & 120263 & 98028 & 105296 & 98028 & 36 & 12 & 57 & 33 & 0 &  - &  - &  - \\
\hline
{ \tiny lighttpd-1913-1914} & 68199 & 48133 & 41374 & 37396 & 49 & 49 & 125 & 125 & 0 &  - &  - &  - \\
\hline
\end{tabular}

\caption{SPR-200-CExt Statistics}
\end{sidewaystable*}

\begin{sidewaystable*}
\small
\centering
\begin{tabular}{|l|c|c|c|c|c|c|c|c|c|c|c|c|}
\hline
\multirow{3}{*}{\bf Defect} & \multicolumn{2}{|c|}{\bf Search Space} & \multicolumn{2}{|c|}{\bf Evaluated} & \multicolumn{2}{|c|}{\bf Plausible} & \multicolumn{2}{|c|}{\bf Plausible} & \multirow{3}{*}{\parbox[c]{1.2cm}{\bf Correct Patches}} & \multicolumn{2}{|c|}{\bf Correct} & {\bf Correct} \\
 &  \multicolumn{2}{|c|}{\bf Templates} & \multicolumn{2}{|c|}{\bf Templates} & \multicolumn{2}{|c|}{\bf Templates} & \multicolumn{2}{|c|}{\bf Patches} &     &  \multicolumn{2}{|c|}{\bf Template Rank} & {\bf Patch Rank} \\
\cline{2-3} \cline{4-5} \cline{6-7} \cline{8-9} \cline{11-12}
& {\bf All} & {\bf Cond.} &  {\bf All} & {\bf Cond.} & {\bf All} & {\bf Cond.} & {\bf All} & {\bf Cond.} &  & {\bf In Space} & {\bf In Plausible} & {\bf in Plausible} \\
\hline
{ \tiny gmp-14166-14167} & 61171 & 14161 & 50781 & 14161 & 33 & 28 & 111 & 106 & 0 &  - &  - &  - \\
\hline
{ \tiny libtiff-5b0217-3dfb33} & 278857 & 65017 & 104524 & 65017 & 1164 & 149 & 1167 & 152 & 1 & 56646 & 209 & 212 \\
\hline
{ \tiny libtiff-d13be7-ccadf4} & 300745 & 228659 & 226189 & 210869 & 1723 & 1723 & 2003 & 1723 & 1 & 6080 & 3 & 3 \\
\hline
{ \tiny lighttpd-2661-2662} & 124449 & 98028 & 108906 & 98028 & 9 & 3 & 10 & 4 & 0 &  - &  - &  - \\
\hline
{ \tiny lighttpd-1913-1914} & 70712 & 48133 & 45056 & 40544 & 55 & 55 & 55 & 55 & 0 &  - &  - &  - \\
\hline
\end{tabular}

\caption{SPR-200-RExt Statistics}
\end{sidewaystable*}

\begin{sidewaystable*}
\small
\centering
\begin{tabular}{|l|c|c|c|c|c|c|c|c|c|c|c|c|}
\hline
\multirow{3}{*}{\bf Defect} & \multicolumn{2}{|c|}{\bf Search Space} & \multicolumn{2}{|c|}{\bf Evaluated} & \multicolumn{2}{|c|}{\bf Plausible} & \multicolumn{2}{|c|}{\bf Plausible} & \multirow{3}{*}{\parbox[c]{1.2cm}{\bf Correct Patches}} & \multicolumn{2}{|c|}{\bf Correct} & {\bf Correct} \\
 &  \multicolumn{2}{|c|}{\bf Templates} & \multicolumn{2}{|c|}{\bf Templates} & \multicolumn{2}{|c|}{\bf Templates} & \multicolumn{2}{|c|}{\bf Patches} &     &  \multicolumn{2}{|c|}{\bf Template Rank} & {\bf Patch Rank} \\
\cline{2-3} \cline{4-5} \cline{6-7} \cline{8-9} \cline{11-12}
& {\bf All} & {\bf Cond.} &  {\bf All} & {\bf Cond.} & {\bf All} & {\bf Cond.} & {\bf All} & {\bf Cond.} &  & {\bf In Space} & {\bf In Plausible} & {\bf in Plausible} \\
\hline
{ \tiny gmp-14166-14167} & 61171 & 14161 & 51587 & 14161 & 29 & 24 & 56 & 51 & 0 &  - &  - &  - \\
\hline
{ \tiny libtiff-5b0217-3dfb33} & 278857 & 65017 & 12964 & 8420 & 111 & 111 & 2604 & 2604 & 0 & 56646 &  - &  - \\
\hline
{ \tiny libtiff-d13be7-ccadf4} & 300745 & 228659 & 13486 & 13486 & 232 & 232 & 4329 & 4329 & 2 & 6080 & 5 & 104 \\
\hline
{ \tiny lighttpd-2661-2662} & 124449 & 98028 & 108266 & 98028 & 10 & 4 & 16 & 10 & 0 &  - &  - &  - \\
\hline
{ \tiny lighttpd-1913-1914} & 70712 & 48133 & 41374 & 37396 & 48 & 48 & 123 & 123 & 0 &  - &  - &  - \\
\hline
\end{tabular}

\caption{SPR-200-RExt-CExt Statistics}
\end{sidewaystable*}

\begin{sidewaystable*}
\small
\centering
\begin{tabular}{|l|c|c|c|c|c|c|c|c|c|c|c|c|}
\hline
\multirow{3}{*}{\bf Defect} & \multicolumn{2}{|c|}{\bf Search Space} & \multicolumn{2}{|c|}{\bf Evaluated} & \multicolumn{2}{|c|}{\bf Plausible} & \multicolumn{2}{|c|}{\bf Plausible} & \multirow{3}{*}{\parbox[c]{1.2cm}{\bf Correct Patches}} & \multicolumn{2}{|c|}{\bf Correct} & {\bf Correct} \\
 &  \multicolumn{2}{|c|}{\bf Templates} & \multicolumn{2}{|c|}{\bf Templates} & \multicolumn{2}{|c|}{\bf Templates} & \multicolumn{2}{|c|}{\bf Patches} &     &  \multicolumn{2}{|c|}{\bf Template Rank} & {\bf Patch Rank} \\
\cline{2-3} \cline{4-5} \cline{6-7} \cline{8-9} \cline{11-12}
& {\bf All} & {\bf Cond.} &  {\bf All} & {\bf Cond.} & {\bf All} & {\bf Cond.} & {\bf All} & {\bf Cond.} &  & {\bf In Space} & {\bf In Plausible} & {\bf in Plausible} \\
\hline
{ \tiny gmp-14166-14167} & 59327 & 22055 & 41316 & 22055 & 34 & 28 & 112 & 106 & 0 &  - &  - &  - \\
\hline
{ \tiny libtiff-5b0217-3dfb33} & 276142 & 85251 & 123778 & 85251 & 237 & 149 & 240 & 152 & 1 & 59068 & 206 & 209 \\
\hline
{ \tiny libtiff-d13be7-ccadf4} & 568172 & 432083 & 250826 & 237142 & 1723 & 1723 & 2003 & 1723 & 1 & 6126 & 3 & 3 \\
\hline
{ \tiny lighttpd-2661-2662} & 191579 & 159862 & 119941 & 113050 & 10 & 5 & 13 & 8 & 0 &  - &  - &  - \\
\hline
{ \tiny lighttpd-1913-1914} & 159739 & 114667 & 37901 & 34739 & 32 & 32 & 32 & 32 & 0 &  - &  - &  - \\
\hline
\end{tabular}

\caption{SPR-300 Statistics}
\end{sidewaystable*}

\begin{sidewaystable*}
\small
\centering
\begin{tabular}{|l|c|c|c|c|c|c|c|c|c|c|c|c|}
\hline
\multirow{3}{*}{\bf Defect} & \multicolumn{2}{|c|}{\bf Search Space} & \multicolumn{2}{|c|}{\bf Evaluated} & \multicolumn{2}{|c|}{\bf Plausible} & \multicolumn{2}{|c|}{\bf Plausible} & \multirow{3}{*}{\parbox[c]{1.2cm}{\bf Correct Patches}} & \multicolumn{2}{|c|}{\bf Correct} & {\bf Correct} \\
 &  \multicolumn{2}{|c|}{\bf Templates} & \multicolumn{2}{|c|}{\bf Templates} & \multicolumn{2}{|c|}{\bf Templates} & \multicolumn{2}{|c|}{\bf Patches} &     &  \multicolumn{2}{|c|}{\bf Template Rank} & {\bf Patch Rank} \\
\cline{2-3} \cline{4-5} \cline{6-7} \cline{8-9} \cline{11-12}
& {\bf All} & {\bf Cond.} &  {\bf All} & {\bf Cond.} & {\bf All} & {\bf Cond.} & {\bf All} & {\bf Cond.} &  & {\bf In Space} & {\bf In Plausible} & {\bf in Plausible} \\
\hline
{ \tiny gmp-14166-14167} & 59327 & 22055 & 42448 & 22055 & 30 & 24 & 57 & 51 & 0 &  - &  - &  - \\
\hline
{ \tiny libtiff-5b0217-3dfb33} & 276142 & 85251 & 16314 & 9388 & 111 & 111 & 2598 & 2598 & 0 & 59068 &  - &  - \\
\hline
{ \tiny libtiff-d13be7-ccadf4} & 568172 & 432083 & 19425 & 19425 & 203 & 203 & 4089 & 4089 & 2 & 6126 & 5 & 104 \\
\hline
{ \tiny lighttpd-2661-2662} & 191579 & 159862 & 116171 & 109330 & 14 & 9 & 34 & 29 & 0 &  - &  - &  - \\
\hline
{ \tiny lighttpd-1913-1914} & 159739 & 114667 & 28702 & 27455 & 24 & 24 & 53 & 53 & 0 & 107499 &  - &  - \\
\hline
\end{tabular}

\caption{SPR-300-CExt Statistics}
\end{sidewaystable*}

\begin{sidewaystable*}
\small
\centering
\begin{tabular}{|l|c|c|c|c|c|c|c|c|c|c|c|c|}
\hline
\multirow{3}{*}{\bf Defect} & \multicolumn{2}{|c|}{\bf Search Space} & \multicolumn{2}{|c|}{\bf Evaluated} & \multicolumn{2}{|c|}{\bf Plausible} & \multicolumn{2}{|c|}{\bf Plausible} & \multirow{3}{*}{\parbox[c]{1.2cm}{\bf Correct Patches}} & \multicolumn{2}{|c|}{\bf Correct} & {\bf Correct} \\
 &  \multicolumn{2}{|c|}{\bf Templates} & \multicolumn{2}{|c|}{\bf Templates} & \multicolumn{2}{|c|}{\bf Templates} & \multicolumn{2}{|c|}{\bf Patches} &     &  \multicolumn{2}{|c|}{\bf Template Rank} & {\bf Patch Rank} \\
\cline{2-3} \cline{4-5} \cline{6-7} \cline{8-9} \cline{11-12}
& {\bf All} & {\bf Cond.} &  {\bf All} & {\bf Cond.} & {\bf All} & {\bf Cond.} & {\bf All} & {\bf Cond.} &  & {\bf In Space} & {\bf In Plausible} & {\bf in Plausible} \\
\hline
{ \tiny gmp-14166-14167} & 95751 & 22055 & 54107 & 22055 & 39 & 28 & 117 & 106 & 1 & 50188 & 32 & 110 \\
\hline
{ \tiny libtiff-5b0217-3dfb33} & 336808 & 85251 & 123207 & 85251 & 1164 & 149 & 1165 & 155 & 1 & 59066 & 209 & 215 \\
\hline
{ \tiny libtiff-d13be7-ccadf4} & 573714 & 432083 & 87760 & 87073 & 309 & 309 & 309 & 309 & 1 & 4629 & 3 & 3 \\
\hline
{ \tiny lighttpd-2661-2662} & 197676 & 159862 & 120127 & 113198 & 8 & 3 & 8 & 3 & 0 &  - &  - &  - \\
\hline
{ \tiny lighttpd-1913-1914} & 168946 & 114667 & 37891 & 34739 & 32 & 32 & 32 & 32 & 0 &  - &  - &  - \\
\hline
\end{tabular}

\caption{SPR-300-RExt Statistics}
\end{sidewaystable*}

\begin{sidewaystable*}
\small
\centering
\begin{tabular}{|l|c|c|c|c|c|c|c|c|c|c|c|c|}
\hline
\multirow{3}{*}{\bf Defect} & \multicolumn{2}{|c|}{\bf Search Space} & \multicolumn{2}{|c|}{\bf Evaluated} & \multicolumn{2}{|c|}{\bf Plausible} & \multicolumn{2}{|c|}{\bf Plausible} & \multirow{3}{*}{\parbox[c]{1.2cm}{\bf Correct Patches}} & \multicolumn{2}{|c|}{\bf Correct} & {\bf Correct} \\
 &  \multicolumn{2}{|c|}{\bf Templates} & \multicolumn{2}{|c|}{\bf Templates} & \multicolumn{2}{|c|}{\bf Templates} & \multicolumn{2}{|c|}{\bf Patches} &     &  \multicolumn{2}{|c|}{\bf Template Rank} & {\bf Patch Rank} \\
\cline{2-3} \cline{4-5} \cline{6-7} \cline{8-9} \cline{11-12}
& {\bf All} & {\bf Cond.} &  {\bf All} & {\bf Cond.} & {\bf All} & {\bf Cond.} & {\bf All} & {\bf Cond.} &  & {\bf In Space} & {\bf In Plausible} & {\bf in Plausible} \\
\hline
{ \tiny gmp-14166-14167} & 95751 & 22055 & 54176 & 22055 & 35 & 24 & 62 & 51 & 1 & 50188 & 28 & 55 \\
\hline
{ \tiny libtiff-5b0217-3dfb33} & 336808 & 85251 & 13784 & 9388 & 111 & 111 & 2604 & 2604 & 0 & 59066 &  - &  - \\
\hline
{ \tiny libtiff-d13be7-ccadf4} & 573714 & 432083 & 19425 & 19425 & 215 & 215 & 4125 & 4125 & 2 & 4629 & 5 & 104 \\
\hline
{ \tiny lighttpd-2661-2662} & 197676 & 159862 & 116171 & 109330 & 16 & 11 & 31 & 26 & 0 &  - &  - &  - \\
\hline
{ \tiny lighttpd-1913-1914} & 168946 & 114667 & 28116 & 26867 & 24 & 24 & 53 & 53 & 0 & 107498 &  - &  - \\
\hline
\end{tabular}

\caption{SPR-300-RExt-CExt Statistics}
\end{sidewaystable*}

\begin{sidewaystable*}
\small
\centering
\begin{tabular}{|l|c|c|c|c|c|c|c|c|c|c|c|c|}
\hline
\multirow{3}{*}{\bf Defect} & \multicolumn{2}{|c|}{\bf Search Space} & \multicolumn{2}{|c|}{\bf Evaluated} & \multicolumn{2}{|c|}{\bf Plausible} & \multicolumn{2}{|c|}{\bf Plausible} & \multirow{3}{*}{\parbox[c]{1.2cm}{\bf Correct Patches}} & \multicolumn{2}{|c|}{\bf Correct} & {\bf Correct} \\
 &  \multicolumn{2}{|c|}{\bf Templates} & \multicolumn{2}{|c|}{\bf Templates} & \multicolumn{2}{|c|}{\bf Templates} & \multicolumn{2}{|c|}{\bf Patches} &     &  \multicolumn{2}{|c|}{\bf Template Rank} & {\bf Patch Rank} \\
\cline{2-3} \cline{4-5} \cline{6-7} \cline{8-9} \cline{11-12}
& {\bf All} & {\bf Cond.} &  {\bf All} & {\bf Cond.} & {\bf All} & {\bf Cond.} & {\bf All} & {\bf Cond.} &  & {\bf In Space} & {\bf In Plausible} & {\bf in Plausible} \\
\hline
{ \tiny gmp-14166-14167} & 448318 & 141182 & 48639 & 40229 & 28 & 28 & 106 & 106 & 0 &  - &  - &  - \\
\hline
{ \tiny libtiff-5b0217-3dfb33} & 1941177 & 1352092 & 376088 & 346011 & 59 & 59 & 59 & 59 & 0 & 309678 &  - &  - \\
\hline
{ \tiny libtiff-d13be7-ccadf4} & 2353240 & 1771135 & 537943 & 510074 & 39 & 39 & 39 & 39 & 1 & 34935 & 3 & 3 \\
\hline
{ \tiny lighttpd-2661-2662} & 1269307 & 1016413 & 158849 & 158849 & 84 & 84 & 108 & 108 & 1 & 73333 & 10 & 18 \\
\hline
{ \tiny lighttpd-1913-1914} & 1183286 & 935121 & 67947 & 67947 & 30 & 30 & 30 & 30 & 0 &  - &  - &  - \\
\hline
\end{tabular}

\caption{SPR-2000 Statistics}
\end{sidewaystable*}

\begin{sidewaystable*}
\small
\centering
\begin{tabular}{|l|c|c|c|c|c|c|c|c|c|c|c|c|}
\hline
\multirow{3}{*}{\bf Defect} & \multicolumn{2}{|c|}{\bf Search Space} & \multicolumn{2}{|c|}{\bf Evaluated} & \multicolumn{2}{|c|}{\bf Plausible} & \multicolumn{2}{|c|}{\bf Plausible} & \multirow{3}{*}{\parbox[c]{1.2cm}{\bf Correct Patches}} & \multicolumn{2}{|c|}{\bf Correct} & {\bf Correct} \\
 &  \multicolumn{2}{|c|}{\bf Templates} & \multicolumn{2}{|c|}{\bf Templates} & \multicolumn{2}{|c|}{\bf Templates} & \multicolumn{2}{|c|}{\bf Patches} &     &  \multicolumn{2}{|c|}{\bf Template Rank} & {\bf Patch Rank} \\
\cline{2-3} \cline{4-5} \cline{6-7} \cline{8-9} \cline{11-12}
& {\bf All} & {\bf Cond.} &  {\bf All} & {\bf Cond.} & {\bf All} & {\bf Cond.} & {\bf All} & {\bf Cond.} &  & {\bf In Space} & {\bf In Plausible} & {\bf in Plausible} \\
\hline
{ \tiny gmp-14166-14167} & 448318 & 141182 & 29653 & 24692 & 24 & 24 & 51 & 51 & 0 &  - &  - &  - \\
\hline
{ \tiny libtiff-5b0217-3dfb33} & 1941177 & 1352092 & 13873 & 13873 & 111 & 111 & 2442 & 2442 & 0 & 309678 &  - &  - \\
\hline
{ \tiny libtiff-d13be7-ccadf4} & 2353240 & 1771135 & 288244 & 284634 & 52 & 52 & 2708 & 2708 & 2 & 34935 & 5 & 104 \\
\hline
{ \tiny lighttpd-2661-2662} & 1269307 & 1016413 & 141022 & 141022 & 117 & 117 & 189 & 189 & 1 & 73333 & 9 & 14 \\
\hline
{ \tiny lighttpd-1913-1914} & 1183286 & 935121 & 60084 & 60084 & 37 & 37 & 84 & 84 & 0 & 297232 &  - &  - \\
\hline
\end{tabular}

\caption{SPR-2000-CExt Statistics}
\end{sidewaystable*}

\begin{sidewaystable*}
\small
\centering
\begin{tabular}{|l|c|c|c|c|c|c|c|c|c|c|c|c|}
\hline
\multirow{3}{*}{\bf Defect} & \multicolumn{2}{|c|}{\bf Search Space} & \multicolumn{2}{|c|}{\bf Evaluated} & \multicolumn{2}{|c|}{\bf Plausible} & \multicolumn{2}{|c|}{\bf Plausible} & \multirow{3}{*}{\parbox[c]{1.2cm}{\bf Correct Patches}} & \multicolumn{2}{|c|}{\bf Correct} & {\bf Correct} \\
 &  \multicolumn{2}{|c|}{\bf Templates} & \multicolumn{2}{|c|}{\bf Templates} & \multicolumn{2}{|c|}{\bf Templates} & \multicolumn{2}{|c|}{\bf Patches} &     &  \multicolumn{2}{|c|}{\bf Template Rank} & {\bf Patch Rank} \\
\cline{2-3} \cline{4-5} \cline{6-7} \cline{8-9} \cline{11-12}
& {\bf All} & {\bf Cond.} &  {\bf All} & {\bf Cond.} & {\bf All} & {\bf Cond.} & {\bf All} & {\bf Cond.} &  & {\bf In Space} & {\bf In Plausible} & {\bf in Plausible} \\
\hline
{ \tiny gmp-14166-14167} & 780425 & 141182 & 27515 & 23574 & 10 & 10 & 29 & 29 & 0 & 214233 &  - &  - \\
\hline
{ \tiny libtiff-5b0217-3dfb33} & 2027218 & 1352092 & 353982 & 324268 & 59 & 59 & 59 & 59 & 0 & 309679 &  - &  - \\
\hline
{ \tiny libtiff-d13be7-ccadf4} & 2414317 & 1771135 & 532541 & 506165 & 39 & 39 & 39 & 39 & 1 & 33699 & 3 & 3 \\
\hline
{ \tiny lighttpd-2661-2662} & 1290186 & 1016413 & 157305 & 157305 & 84 & 84 & 105 & 105 & 1 & 73333 & 10 & 18 \\
\hline
{ \tiny lighttpd-1913-1914} & 1206699 & 935121 & 67947 & 67947 & 31 & 31 & 31 & 31 & 0 &  - &  - &  - \\
\hline
\end{tabular}

\caption{SPR-2000-RExt Statistics}
\end{sidewaystable*}

\begin{sidewaystable*}
\small
\centering
\begin{tabular}{|l|c|c|c|c|c|c|c|c|c|c|c|c|}
\hline
\multirow{3}{*}{\bf Defect} & \multicolumn{2}{|c|}{\bf Search Space} & \multicolumn{2}{|c|}{\bf Evaluated} & \multicolumn{2}{|c|}{\bf Plausible} & \multicolumn{2}{|c|}{\bf Plausible} & \multirow{3}{*}{\parbox[c]{1.2cm}{\bf Correct Patches}} & \multicolumn{2}{|c|}{\bf Correct} & {\bf Correct} \\
 &  \multicolumn{2}{|c|}{\bf Templates} & \multicolumn{2}{|c|}{\bf Templates} & \multicolumn{2}{|c|}{\bf Templates} & \multicolumn{2}{|c|}{\bf Patches} &     &  \multicolumn{2}{|c|}{\bf Template Rank} & {\bf Patch Rank} \\
\cline{2-3} \cline{4-5} \cline{6-7} \cline{8-9} \cline{11-12}
& {\bf All} & {\bf Cond.} &  {\bf All} & {\bf Cond.} & {\bf All} & {\bf Cond.} & {\bf All} & {\bf Cond.} &  & {\bf In Space} & {\bf In Plausible} & {\bf in Plausible} \\
\hline
{ \tiny gmp-14166-14167} & 780425 & 141182 & 27505 & 23574 & 6 & 6 & 13 & 13 & 0 & 214233 &  - &  - \\
\hline
{ \tiny libtiff-5b0217-3dfb33} & 2027218 & 1352092 & 37909 & 37909 & 111 & 111 & 2264 & 2264 & 0 & 309679 &  - &  - \\
\hline
{ \tiny libtiff-d13be7-ccadf4} & 2414317 & 1771135 & 285546 & 283944 & 52 & 52 & 2708 & 2708 & 2 & 33699 & 5 & 104 \\
\hline
{ \tiny lighttpd-2661-2662} & 1290186 & 1016413 & 136816 & 136816 & 117 & 117 & 189 & 189 & 1 & 73333 & 9 & 14 \\
\hline
{ \tiny lighttpd-1913-1914} & 1206699 & 935121 & 64918 & 64918 & 28 & 28 & 65 & 65 & 0 & 297229 &  - &  - \\
\hline
\end{tabular}

\caption{SPR-2000-RExt-CExt Statistics}
\label{tab:app-last}
\end{sidewaystable*}

%%% Local Variables:
%%% TeX-master: "paper"
%%% End:

%\begin{thebibliography}{}

%\bibitem{smith02}
%Smith, P. Q. reference text

%\end{thebibliography}

\end{document}